\documentclass[lettersize,journal]{IEEEtran}

\usepackage{multirow}
\usepackage{graphicx}
\usepackage{amsmath}
\usepackage{color}
\usepackage{epsfig}
\usepackage{amsfonts}
\usepackage{amssymb}
\usepackage{amsthm}
\usepackage{epstopdf}
\usepackage{caption}
\usepackage{subcaption}
\usepackage{mathtools}
\usepackage{url}
\def\BibTeX{{\rm B\kern-.05em{\sc i\kern-.025em b}\kern-.08em
		T\kern-.1667em\lower.7ex\hbox{E}\kern-.125emX}}
\usepackage{balance}
\usepackage{bm}
\usepackage{hyperref}
\usepackage{makecell}
\usepackage{mathtools}
\usepackage{epstopdf}
\usepackage{cite}
\usepackage{float}
\usepackage{threeparttable}

\usepackage[ruled,vlined,linesnumbered,commentsnumbered]{algorithm2e}
\usepackage{setspace}

\usepackage{soul}   
\setstcolor{red}
\setstcolor{violet}
\usepackage{xcolor} 

\newcommand{\mbf}[1]{\mathbf{#1}}
\newcommand{\bols}[1]{\boldsymbol{#1}}
\newcommand{\tbf}[1]{\textbf{#1}}
\newcommand{\mbb}[1]{\mathbb{#1}}
\newcommand{\mrm}[1]{\mathrm{#1}}

\newcommand\Tau{\mathcal{T}}

\usepackage{setspace}

\begin{document}
	\title{{\huge Sum-Rate Maximization {of} RIS-Aided {Digital and Holographic Beamformers in} MU-MISO Systems}}

	\author{
		\IEEEauthorblockN{Pavan Kumar Gadamsetty\IEEEauthorrefmark{1}, K. V. S. Hari\IEEEauthorrefmark{1} and Lajos Hanzo\IEEEauthorrefmark{2}} \\
		\IEEEauthorblockA{\IEEEauthorrefmark{1}Department of Electrical Communication Engineering, Indian Institute of Science, Bengaluru 560012, India}
		\IEEEauthorblockA{\IEEEauthorrefmark{2}School of Electronics and Computer Science, University of Southampton, SO17 1BJ Southampton, U.K.}
	}
	
	\maketitle

	\begin{abstract}
		
		Reconfigurable holographic surfaces (RHS) {are intrinsically amalgamated} with reconfigurable intelligent surfaces (RIS), {for beneficially ameliorating} the signal propagation environment. {This potent architecture} significantly improves the system performance {in} non-line-of-sight {scenarios} at a {low} power consumption. {Briefly,} the RHS technology integrates ultra-thin, lightweight antennas onto the transceiver, 
		{for creating sharp, high-gain} directional beams. 
		{We formulate a user} sum-rate maximization problem for {our} RHS-RIS-based hybrid {beamformer}. {Explicitly, we} jointly design the digital, holographic, and passive beamformers {for maximizing} the sum-rate of all {user equipment} (UE). To tackle the {resultant} nonconvex optimization problem, we propose an alternating maximization (AM) framework {for decoupling} and iteratively {solving} the subproblems involved. Specifically, we employ the zero-forcing criterion for the digital {beamformer}, leverage fractional programming to determine the radiation amplitudes of the RHS and utilize the Riemannian conjugate gradient algorithm {for  optimizing the} RIS phase shift matrix of the passive {beamformer}. Our simulation results demonstrate that the proposed RHS-RIS-based hybrid beamformer outperforms {its conventional counterpart operating} without an RIS in multi-UE scenarios. {The sum-rate improvement attained ranges from 8 bps/Hz to 13 bps/Hz for various transmit powers at the base station (BS) and at the UEs, which is significant.}
		
	\end{abstract}


	\IEEEpeerreviewmaketitle

	\section{Introduction}

	The next-generation wireless communication systems primarily {aim for} enhancing {the} reliable data transfer speeds, {for} reduced latency, and {for} providing ubiquitous connectivity~\cite{Harald_2023_IEEECOMSurvey_6Groad_Vision,Vucetic_2019_IEEEJIOT_Low_Latency}. In particular, the millimeter-wave (mmWave) band, has emerged as one of the leading candidates for spectrum exploitation to address the {impending} spectrum scarcity and {to} facilitate high-speed data delivery. This mmWave band presents clear advantages owing to the rapid {advances} in its sophisticated circuit design~\cite{Gutierrez_2011_IEEEProc_mmwave_circuit}. In conjunction with the existing   mmWave~\cite{Hanzo_2017_IEEE_Comsurvey_mmwave,Lajos_2024_IEEETVT_mmwave_RIS_impair,Chau_2024_IEEEWCL_Far_near} bands,
    reconfigurable intelligent surfaces (RIS) {have} been proposed {for enhancing} the performance of future wireless systems~\cite{Lajos_2024_IEEEVTMag_RIS_6G,Chau_2019_IEEETWC_RIS,Hanzo_2023_IEEETCOM_RIS_capacity}. {Briefly, a} RIS consists of a metasurface {having} programmable reflecting elements (PREs) that passively manipulate {the} incident waves, directing them towards desired destinations, unlike traditional signal relaying methods~\cite{Sergei_2023_IEEEAnt_reflectarrays}.
	Thus, {RIS} address the limitations of conventional wireless channels by {harnessing} the unique {capability} of metasurfaces to manipulate {the} electromagnetic waves~\cite{Zhang_2020_IEEE_Commagazine_RIS_appli}. This manipulation enables various applications, including arbitrary aperture beamforming~\cite{Zhang_2020_IEEE_Commagazine_RIS_appli}, polarization conversion~\cite{Tong_2013_IEEE_TAP_linearpolari}, and beam focusing~\cite{Amir_2019_ASIACONF_Beamfocus}.  
	
    Wireless signals in the mmWave bands experience significant path loss, resulting in {potential} performance degradation. To address this issue in  mmWave systems, it is essential to jointly design the active beamformer at the base station (BS) and the passive beamformer at the RIS. {As} for {the} active beamformer at the BS, {originally} analog beamforming  was proposed, which is based on a fully-connected (FC) architecture, where each radio frequency (RF) chain was connected to all antennas~\cite{Zhang_2018_IEEETV_ABF}. However, this approach required an excessive number of phase shifters, even for a low number of RF chains, leading to considerable power consumption.
	Hybrid beamforming (HBF) emerged as a practical low-power solution, combining analog and digital (baseband) beamformers, where each RF chain is connected to a subset of antennas~\cite{Qian_2021_IEEETVT_HBF_FC_SC_perf}.

	
	\begin{table*}[t]
		\centering
		
		\caption{\label{tab:References_check} {COMPARING OUR CONTRIBUTION TO THE EXISTING LITERATURE}}
		\label{tab:widetable}
		\begin{tabular}{|l|l|l|l|l|l|l|l|l|l|}
			\hline 
            &{~\cite{Khaled_2016_IEEEJSTSP_mmWave_Alternating}} & {~\cite{Erik_2020_IEEETWC_RIS_weightedsumrate}} &{~\cite{Kit_2021_IEEETCOM_Wtedsumsecrecyrate}}
			&{~\cite{Lajos_2022_IEEETVT_RIS_ND}} 
			&{~\cite{Lajos_2023_IEEEOJCom_RIS_ND_ADMM}}
			&{~\cite{Lingyang_2022_IEEETWC_RHS_wireless}} 
			&{~\cite{Cheng_2023_IEEECOML_HBF_MMSE}} 
			&{~\cite{Zhu_2023_IEEETWC_HBF_jointprecode}}
			& \tbf{Proposed}\\ \hline
			mmWave channel &  \checkmark &  &\checkmark & &\checkmark &\checkmark  &\checkmark & \checkmark&\checkmark\\ \hline
			Multi-UE &   &\checkmark &\checkmark &\checkmark & \checkmark&\checkmark  &\checkmark &\checkmark&\checkmark\\ \hline
			Optimal power allocation  &  &\checkmark  & \checkmark& \checkmark&\checkmark & \checkmark  &\checkmark &&\checkmark \\ \hline
			Fractional programming &  & \checkmark & & & &  & &\checkmark&\checkmark\\ \hline

			Manifold optimization& \checkmark & \checkmark & & & &  & &&\checkmark\\ \hline
     		RIS &   &\checkmark &\checkmark & \checkmark&\checkmark &  & & &\checkmark\\ \hline
			Holographic beamformer &  &  & & & &\checkmark  & &&\checkmark\\ \hline
			Sum-rate &  &\checkmark  &\checkmark & &\checkmark &\checkmark  & & \checkmark&\checkmark\\ \hline
			Dinkelbach-based method &  &  & & & &  & & &\checkmark\\ \hline
			RHS-RIS combination &  &  & & & &  & & &\checkmark\\ \hline
			{Mutual coupling at RHS} &  &  & & & &  & & &\checkmark\\ \hline

		\end{tabular}
		\begin{tablenotes}[para] \footnotesize
			
		\end{tablenotes}
		
	\end{table*} 	
	

 Extensive research has been focused on the design of HBF, with multiple studies investigating {the challenging} aspects of this design~\cite{Hanzo_2022_IEEETVT_HBF_latency,Li_2022_IEEETCOM_HBF_precoding,Ottersten_2022_IEEETWC_HBF_Leo,Hanzo_2022_IEEEIOTJ_HBF_RIS_Subarrays,Qian_2021_IEEETVT_HBF_FC_SC_perf,Cheng_2023_IEEECOML_HBF_MMSE,Zhu_2023_IEEETWC_HBF_jointprecode}, {including} multiuser (MU) HBF~\cite{Hanzo_2022_IEEEIOTJ_HBF_RIS_Subarrays,Cheng_2023_IEEECOML_HBF_MMSE}. {A} predominant focus of MU beamforming is the maximization of sum-rate (SR). In~\cite{Wang_2022_IEEETVT_sumrate}, the authors proposed a joint design of PREs and MU beamforming
 in a narrow-band scenario to maximize {the} users' sum-rate.
 Additionally, weighted sum-rate (WSR) algorithms have been proposed in a Multiple-Input Single-Output (MISO) multiuser downlink {scenario} in~\cite{Erik_2020_IEEETWC_RIS_weightedsumrate,Kit_2021_IEEETCOM_Wtedsumsecrecyrate}. These algorithms {jointly} optimize beamforming at the BS and the phase coefficients of RIS elements to enhance {the} WSR. Furthermore, alternating optimization methods have been proposed {for maximizing the} achievable rate by jointly optimizing {the} transmit {beamformer} and the non-diagonal RIS phase shift matrix in~\cite{Lajos_2022_IEEETVT_RIS_ND, Lajos_2023_IEEEOJCom_RIS_ND_ADMM}. In the existing HBF techniques, phased arrays are used at the BS, with a RIS positioned between the BS and the user equipments (UEs) {for preventing line-of-sight (LOS) blocking}.

  However, even when using phased arrays-based HBF in the mmWave band, the associated hardware costs and power consumption {remain excessive, imposing} a significant challenge~\cite{David_2015_IEEEProc_Phasedarrayevolution}. {This is because phased arrays require numerous phase shifters and power amplifiers to construct phase-shifting circuits for accurate beamforming.} Additionally, as the {operating} frequency of massive multiple-input multiple-output (MIMO) systems increases, the implementation of phased arrays becomes prohibitive, severely hindering their future development. Therefore, there is an urgent need for developing {bespoke} technologies to meet the exponentially increasing data demands in {next-generation} wireless communications.

  {In contrast to RIS}, the reconfigurable holographic surfaces (RHS) {are capable of addressing} the limitations of existing antenna technologies~\cite{Lingyang_2022_IEEETWC_RHS_wireless,Hanzo_2023_IEEECommLet_Holo_tutorial,Song_2022_IEEEJSAC_holographic_Leo,Song_2022_IEEEJSAC_holographic_ISAC}, {such as conventional} phased array architectures. RHS, proposed as a representative metamaterial antenna~\cite{David_2016_IEEEAntennaLetter_RHS_wave}, leverages the holographic interference principle to control the radiation amplitude of incoming electromagnetic waves. For {supporting} flexible beam steering, the RHS {relies on} a large number of metamaterial radiation elements connected {to} RF chains and {it} is generally integrated with the transceivers. This approach enables the creation of large arrays, while maintaining compact and lightweight transceiver hardware~\cite{Lingyang_2021_IEEE_RHS_intro}.

{The physical structure of RHSs {is} different from that of the reconfigurable intelligent surfaces (RISs)~\cite{Lajos_2024_IEEEVTMag_RIS_6G}. Specifically, an RHS integrates its RF front end into a PCB, allowing for convenient transceiver implementation without requiring an extra control link to construct the holographic pattern. By contrast, an RIS places its RF front end on the outside in support of its reflective action, hence necessitating an additional control link between the RIS and the transmitter to adjust the phase shifts and/or radiation amplitudes. Because of these structural differences, RHSs typically serve as transmit and receive antennas, while RISs are commonly used as relays. Consequently, the active and passive beamforming design of RHSs and RISs are quite different.}

{For improving the system's sum-rate, it is beneficial to simultaneously harness the advantages of both an RHS and RIS. This joint deployment is particularly useful in dense environments exhibiting blocked line-of-sight paths between BS and UEs. Passive RIS beamforming enhances the received signals, while RHS transmitters promptly adapt to fast-fading channels. Despite the challenge of jointly designing the passive and holographic beamformers, integrating an RHS and RIS significantly improves the signal quality, beamforming, energy efficiency, coverage, and cost-effectiveness. This potent combination promises to beneficially ameliorate the signal propagation, enhance coverage and increase the capacity, hence paving the way for advanced wireless communication networks.}

The authors {of}~\cite{Lingyang_2022_IEEETWC_RHS_wireless} introduced the RHS in a mmWave system and solved a sum-rate maximization problem for RHS-based HBF in the absence of a RIS. However, even with improved beam control, RHS-aided systems yielded poor performance in the absence of a stable line-of-sight link, thereby emphasizing the need to employ an RIS~\cite{Alouini_2021_IEETCOM_RHS_RIS_terahertz}.
In~\cite{Bhavani_2023_IEEEICASSP_RIS_RHS_DFRC}, the authors exploit both RHS and RIS systems in the context of a Dual-function radar-communications (DFRC) system to maximize the radar signal-to-interference-plus-noise ratio (SINR), while ensuring {the required} communication SINR {for} all UEs. None of the existing works utilized both RHS and RIS technologies and solved the problem of obtaining the sum-rate expression for a mmWave multi-UE system.

Against the above background, this is the first piece of work
{maximizing} the sum-rate through the joint optimization of digital, holographic, and passive beamformers.
The {resultant} optimization problem is nonconvex {along} with coupled variables. To address this challenge, we decouple the problem into several subproblems, which are then solved using an alternating maximization (AM) algorithm~\cite{Bin_2021_IEEETVT_AMalgo}. Specifically, we employ zero-forcing beamforming~\cite{Shlomo_2008_IEEETSP_ZFBF} for the digital beamforming, utilize fractional programming~\cite{Wei_2018_IEEETSP_fractionalprogram} to determine the radiation amplitudes of the RHS elements, and leverage the Riemannian conjugate gradient algorithm~\cite{Rodolphe_2014_JournML_manopt} to obtain the optimal RIS phase shift matrix for the passive {beamformer}. 
Simulation results demonstrate the effectiveness of the proposed algorithm, exhibiting improvements over existing RHS-based {methods operating} without an RIS.

	\noindent {We summarize the main contributions of the paper as}:
	
	\begin{enumerate}
		
		\item A RHS-RIS-aided MU mmWave MISO system model is
		developed, {for supporting} single-antenna UEs. The end-to-end channel gains are determined for this system, considering scenarios {both} with and without the RIS.  Subsequently, a sum-rate maximization problem is formulated to determine the joint active and passive beamformers for the RIS system with the objective of maximizing the system's sum-rate. {Again,} this optimization problem is non-convex.

		\item We proposed an alternating maximization algorithm for HBF design, aiming {for maximizing} the sum-rate of the UEs. The front-end digital beamformer matrix is obtained via {the} zero-forcing technique, followed by optimal power allocation. The holographic beamforming subproblem is solved using fractional programming. The unit modulus constraints of the RIS phase shift matrix define a Riemannian manifold, leading us to propose a manifold optimization-based algorithm.
		
		\item Extensive simulations demonstrate the remarkable benefits of the proposed algorithm {in terms of the signal-to-interference-plus-noise ratio (SINR)} and the number of UEs in the system. The numerical results indicate a substantial sum-rate {improvement}.

	\end{enumerate}

	The remainder of this paper is organized as follows. Section~II presents the system model. In Section~III, the proposed AM algorithm is discussed. {The performance} of the proposed algorithm is discussed in Section~IV {and we conclude in} Section~V.
	
 Notation: Scalars are denoted by italic letters, lower (upper) {case} bold letters denote column vectors (matrices). {The} super-scripts ${(.)^{-1}}, {(.)^{*}}, {(.)^H}$ represent inverse, complex conjugate, and Hermitian operators, respectively. The symbol $|.|_F$ denotes the Frobenius norm of a matrix, {the} $\circ$ operator denotes the element-wise product, {while} $\Re{.}$ and $\Im{.}$ represent the real and imaginary part, respectively. For a matrix $\mbf{A}$, $\mbf{A}(i,i)$ represents the $i$-th diagonal element, and Tr($\mbf{A}$) denotes its trace. Furthermore, $\mbf{A}_{i,j}$ denotes the element in the $i$-th row and the $j$-th column. For a vector $\mbf{a}$, $\mbf{a}(i)$ represents the $i$-th element, diag($\mbf{a}$) denotes the diagonal matrix whose diagonal elements are the corresponding elements in $\mbf{a}$, and $\bols{1}_K$ is an all-one vector of dimension $K\times1$.


	\section{System Model}
	
	In this section, we will first present the system model and
	channel model of the mmWave MISO system {considered}.
	{Then we} formulate the RHS-RIS-aided HBF problem.\footnote{{In this work we have considered switch-controlled RHS-aided beamforming architecture~\cite{chau_2023_IEEECOMSurvey_holo,Lajos_2024_IEEETCOM_holo,Lingyang_2022_IEEETWC_RHS_wireless}}.}

\begin{figure*}[hbt!]
		\centering
		\includegraphics[width=0.75\linewidth]{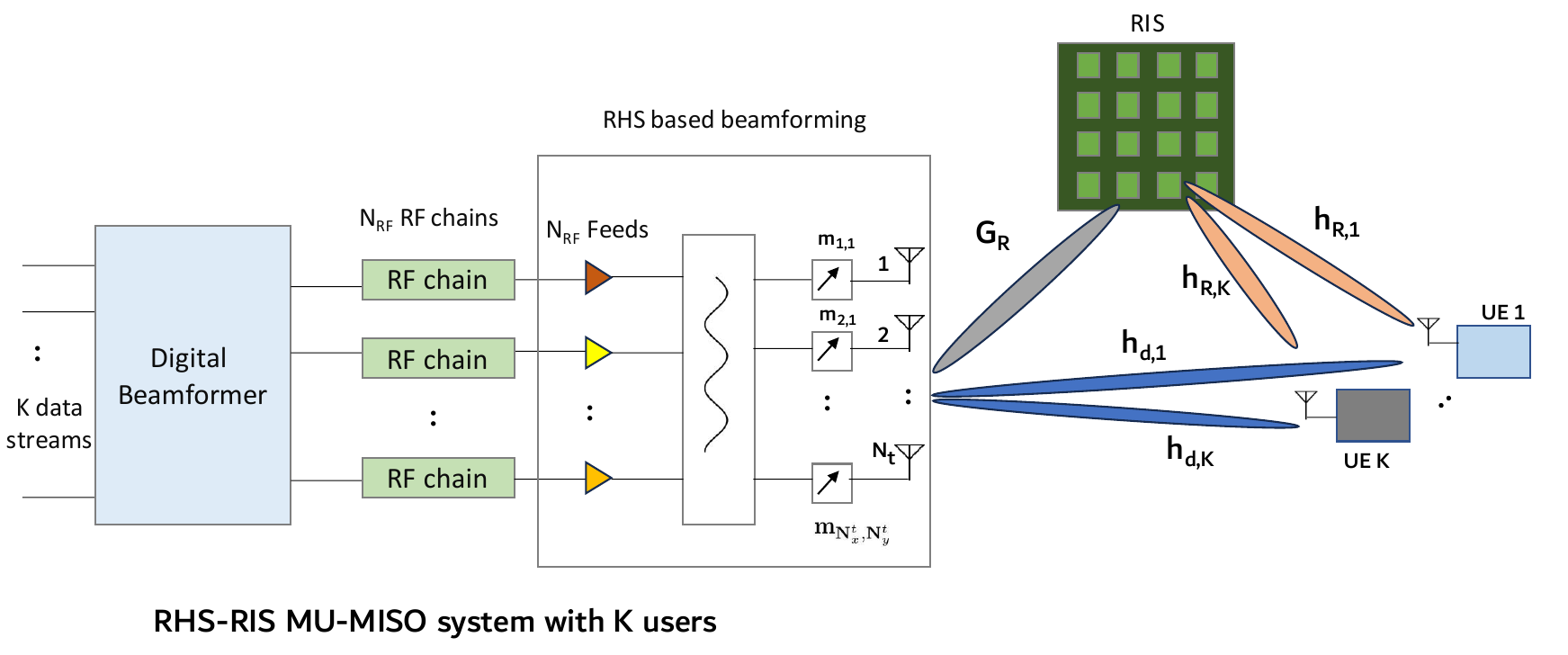}
		\caption{\small Overview of the RHS-RIS MU-MISO system having  $K$ UEs with one receive antenna, a BS with $N_t$ RHS elements and an RIS with $N_{RIS}$ elements.}
		\label{Fig:RHS_RIS_system}
	\end{figure*}

	Consider {the} mmWave multi-UE MISO {downlink (DL)} {scenario}, as illustrated in Fig. \ref{Fig:RHS_RIS_system}. The system consists of a uniform planar array (UPA) {having} $N_t$ {RHS elements} at the BS, an RIS featuring $N_{RIS}$ elements, and $K$ UEs {having a single} receive antenna {(RA)} each. The antennas at the BS (RIS) along the $x$- and $y$-axes are denoted as $N_{t}^x (N^x_{RIS})$ and $N_{t}^y (N^y_{RIS})$, respectively. The BS transmits $K$ data streams, and each UE {receives} a single data stream from the BS. To enhance {the} beamforming capabilities, a RHS is employed at the BS, utilizing holographic techniques. The RHS is fed by $N_{RF}$ radio-frequency (RF) chains, eliminating the need for phase-shifters.

	Let $\mbf{s} \in \mbb{C}^{K\times1}$ {represent} the transmitted symbol before beamforming at the BS, where $\mbf{s} = [s_1, \ldots, s_K]^T$, and $s_k$ denotes the information signal of the $k$-th UE, for $k = 1,\ldots,K$. The symbols $s_k$ are assumed to be independent with an average power of one, i.e., $\mbb{E}[\mbf{s}\mbf{s}^H] = \mbf{I}_K$. The symbol vector $\mbf{s}$ is first precoded by the digital beamformer matrix $\mbf{F} = [\mbf{f}_1, \ldots, \mbf{f}_K] \in \mbb{C}^{N_{RF} \times K}$ and subsequently passed through the RF chains to the RHS beamformer $\mbf{M}_v \in \mbb{C}^{N_t \times N_{RF}}$. The RIS is assumed to have a diagonal phase shift matrix $\bols{\Theta}_{RIS} =$ diag$\{e^{j\theta_{1}}, \hdots,e^{j\theta_{N_{RIS}}}\} \in \mbb{C}^{N_{RIS}\times N_{RIS}}$ with $N_{RIS}$ non-zero diagonal entries.

	The matrices $\mbf{H}_{d} = [\mbf{h}_{d,1},\hdots,\mbf{h}_{d,K}]^H \in \mathbb{C}^{K \times N_t}$, $\mbf{H}_R = [\mbf{h}_{R,1},\hdots,\mbf{h}_{R,K}]^H  \in \mbb{C}^{K\times N_{RIS}}$
	and $\mbf{G}_{R} = [\mbf{g}_{R,1},\hdots,\mbf{g}_{R,N_{t}} ] \in \mathbb{C}^{N_{RIS} \times N_{t}}$ represent the direct channel {spanning} from the BS to $K$ UEs, from the RIS to $K$ UEs, and from the BS to RIS, respectively. Therefore, the effective channel emerging from
	the BS to the $K$ UEs can be expressed as $\mbf{H}_{tot} =
	\mbf{H}_{d} + \mbf{H}_{R}\bols{\Theta}_{RIS} \mbf{G}_R$. In other words, {we have} $\mbf{H}_{tot} = [\mbf{h}_{tot,1},\hdots,\mbf{h}_{tot,K}]^H$, where
	$\mbf{h}_{tot,k}^H \in \mbb{C}^{1\times N_t}$
	$(k = 1, 2,\hdots, K)$ represents the complete channel vector
	from the BS to the $k$-th single-antenna UE.

\subsection{Channel model}
{A realistic channel model should account for both the large-scale fading and small-scale fading characteristics. However, the fading channel models commonly used in traditional MISO systems may not accurately represent the characteristics of mmWave channels. This discrepancy arises from the significant free-space path loss and the presence of large, closely-packed antenna arrays in mmWave environments. Therefore, to accurately capture the characteristics of mmWave propagation, the Saleh-Valenzuela channel model~\cite{Hanzo_2017_IEEE_Comsurvey_mmwave,Hongbin_2020_IEEETWC_mmwave_channel} is utilized. The mmWave channel can be represented as 
\begin{align} 
	&\mathbf {h}_{d,k}^H=\sqrt{\frac{N_t}{L_d}}\sum _{l_d=1}^{L_{d}}\alpha _{l_d}^{}\mathbf {a}_{r}(\phi _{l_d}^{r})\mathbf {a}_{t}(\phi _{l_d}^{t}, \varphi _{l_d}^{t})^{H},  \\
	&{\mathbf {h}}_{R,k}^H=\sqrt{\frac{N_{RIS}}{L_{ru}}}\sum _{l_u=1}^{L_{ru}}\alpha _{l_u}^{}\mathbf {a}_{r}(\phi _{l_u}^{r})\mathbf {a}_{t}(\phi _{l_u}^{t}, \varphi _{l_u}^{t})^{H},  \\
	&{\mathbf{G}}_{R}=\sqrt{\frac{N_{RIS}N_t}{L_{br}}}\sum _{l_b=1}^{L_{br}}\alpha _{l_b}^{}\mathbf {a}_{r}(\phi _{l_b}^{r}, \varphi _{l_b}^{r})\mathbf {a}_{t}(\phi _{l_b}^{t}, \varphi _{l_b}^{t})^{H},
	\label{Eq:mmwave_chan_model}
\end{align} 
	where $L_{d}, L_{ru}$ and $L_{br}$ denote the number of
	multipath components in $\mbf{h}_{d,k},{\mbf{h}}_{R,k},$ and ${\mbf{G}}_{R}$, respectively. Let the first path in ${\mathbf {h}}_{R,k}$ and ${\mathbf{G}}_{R}$ denote the
	line-of-sight components. Furthermore, $\alpha_{l_*}$ denotes the complex channel gain of the ${l_*}$-th path, while $(\phi^r_{*},\varphi^r_{*})$ and $(\phi^t_{*},\varphi^t_{*})$ 
	refer to the physical angle of arrival and angle of departure, respectively. The vectors $\mbf{a}_r(\phi^r_{*},\varphi^r_{*})$ and $\mbf{a}_t(\phi^t_{*},\varphi^t_{*})$ represent the antenna array response vectors. It is assumed that each UE has a single receive antenna, and thus, we have $\mathbf {a}_{r}(\phi _{l_u}^{r})$ = $\mathbf {a}_{r}(\phi _{l_d}^{r})$ = ${1}$.  
	The array response vectors can be written as	
	\begin{align} 
		\mathbf {a}_{z}\left ({\phi, \varphi }\right)= &\sqrt \frac{1}{N^xN^y}\Biggl [1, \ldots, e^{j \frac {2 \pi }{\lambda } d \left ({{n^x \sin \phi \sin \varphi +n^y  \cos \varphi }}\right)},\ldots, \nonumber \\& e^{j \frac {2 \pi }{\lambda } d \left ({{\left ({N^{x}-1}\right) \sin \phi \sin \varphi+\left ({N^{y}-1}\right)  \cos \varphi }}\right)}\Biggr]^{T}. 
	\end{align} }
Here, $z \in \{r, t\}$ and, $0 \leq n^x \leq (N^x-1)$ and $0 \leq n^y \leq (N^y-1)$. The {variables} $N^x$ and $N^y$ represent the number of horizontal and vertical elements, respectively, of the UPA in the 2D plane. 
{The inter-element spacing along the $x$- and $y$-axes for the UPA at the BS and the RIS is denoted by $d$, which is dependent on the wavelength $\lambda$.}
Additionally, ${\phi} \in [0, 2\pi]$ and ${\varphi} \in [0, \pi/2]$ represent the azimuth and elevation angles, respectively.


\subsection{RHS Transmitter}

The RHS lacks digital processing capability, {requiring} the BS to perform signal processing at the baseband. As shown in Fig. \ref{Fig:RHS_RIS_system}, the BS encodes $K$ distinct data streams using a digital beamformer matrix $\mathbf{F}$ and subsequently up-converts the processed signals to the carrier frequency through RF chains. Each RF chain is linked to a feed on the RHS, aligning the number of RF chains with the number of feeds $N_{RF}$. Every RF chain transmits the up-converted signals to its connected feed. The feed then converts the high-frequency current into an electromagnetic wave (reference wave) propagating on the RHS. Since the feeds of the RHS directly connect with the RF chains, there is no channel or attenuation between the BS and the RHS~\cite{Lingyang_2022_IEEETWC_RHS_wireless}. To generate the desired beams, the radiation amplitude of the reference wave at each element is controlled by a holographic beamformer {matrix} ${\mbf{M}}$.

The RHS has $N_{RF}$ feeds and $({N_t^{x}}\times {N_t^{y}})$ discrete elements. The electromagnetic response is given by
\begin{equation}
	{\mbf{M}_v} = {\mbf{M}}{{\mathbf{V}}} \in {\mbb{C}^{N_t^xN_t^y \times {N_{RF}}}},
\end{equation}
\noindent where each element of the matrix $\mbf{V}$ {is} represented as $\mbf{V}(p,q) = e^{-2\pi\gamma D_{p,q}/\lambda}$, where $D_{p,q}$ denotes the distance between the $p$-th RHS element and the $q$-th feed. The matrix $\mbf{M}=\mrm{diag}[m_{1,1},\ldots,m_{1,N_t^y},\ldots,m_{N_t^x,1},\ldots,m_{N_t^x,N_t^y}]$ is a diagonal matrix, with amplitude-control beamformer values $0 \leq m_{x,y} \leq 1$ for each $(x,y)$-th RHS element. \footnote{{We use the terms BS antennas and RHS elements interchangeably.}}

The dimensions of the holographic surface are represented by $N_t^x$ and $N_t^y$. The parameter $\gamma$ {denotes} the refractive index of the material on the RHS. Thus, when the structure of the holographic surface is fixed, the matrix $\mathbf{V}$ is pre-determined.


\subsubsection{Holographic interference principle}
\noindent
{An RHS functions as a leaky-wave antenna comprising three layers, i.e., $N_{RF}$ feeds connect with the corresponding RF chain embedded in the lowest layer. The waveguide structure in the middle layer guides the reference wave to the $N_t$ discrete sub-wavelength metamaterial elements integrated on the top layer~\cite{Shankar_2024_IEEETAES_RIS_RHS_DFRC}. Specifically, at the $(x,y)$-th RHS element, the
	reference wave imported from the $q$-th feed and the desired wave propagating in the target direction $(\phi_0,\varphi_0)$ are characterized by the following expressions, respectively:}
\begin{align}
	\Psi_{obj}(\mbf{r}_{x,y},\phi_0,\varphi_0) &= \textrm{exp}(-j\mbf{k}_f(\phi_0,\varphi_0) \cdot  \mbf{r}_{x,y}), \\
	\Psi_{ref}(\mbf{r}_{x,y}^q) &= \textrm{exp}(-j\mbf{k}_r \cdot \mbf{r}_{x,y}^{q}).
	\label{Eq:holo_1}
\end{align}

{Here, $\mbf{k}_f$ represents the desired directional channel vector in free space, $\mbf{k}_r$ is the channel vector of the reference wave, $\mbf{r}_{x,y}$ is the position vector of the $(x, y)$-th radiation element, and $\mbf{r}_{x,y}^q$ is the distance vector of the link between the feed $q$ and $(x, y)$-th radiation element. The interference between the reference wave and the desired object wave is defined as follows:}
\begin{equation}
	\Psi_{intf}(\mbf{r}_{x,y}^{q},\phi_0,\varphi_0) = \Psi_{obj}(\mbf{r}_{x,y}^{},\phi_0,\varphi_0) \Psi^*_{ref}(\mbf{r}_{x,y}^{q}).
	\label{Eq:holo_2}
\end{equation}

{The RHS can only generate the holographic pattern using the fixed reference wave given in~\eqref{Eq:holo_2}. When the holographic pattern is excited by the reference wave, we have the following result:}
\begin{align}
	\Psi_{intf}(\mbf{r}_{x,y}^{q},\phi_0,\varphi_0) &\Psi_{ref}(\mbf{r}_{x,y}^{q})  \propto \nonumber \\& \Psi_{obj}(\mbf{r}_{x,y}^{},\phi_0,\varphi_0) |\Psi_{ref}(\mbf{r}_{x,y}^{q}) |^2.
	\label{Eq:holo_3}
\end{align}

{To generate the desired wave in the direction $(\phi_0,\varphi_0)$ we control the interference as described in~\eqref{Eq:holo_2}. Furthermore, it is clear from the~\eqref{Eq:holo_2} and~\eqref{Eq:holo_3} that generating the interference waves requires phase adjustment.
However, in contrast to conventional phased arrays, the elements of a RHS can only adjust the radiation amplitude of the reference wave.} {Each radiation element is electrically tuned to resonate at a specific frequency and to emit a reference wave. The specific elements whose emitted waves are aligned in phase with the desired directional beam (i.e., the sum of all radiation elements' waves) are tuned to emit strongly, while those that are out of phase are adjusted to radiate weakly or not at all~\cite{Lingyang_2021_IEEE_RHS_intro}. The real part of the interference (i.e., Re$[\Psi_{intf}]$) represents the cosine of the phase difference between the object wave and the reference wave. As this phase difference increases, Re$[\Psi_{intf}]$ decreases, meeting the requirement for amplitude control. Thus, Re$[\Psi_{intf}]$ serves as a measure of the radiation amplitude for each radiation element. To ensure non-negative values,  Re$[\Psi_{intf}]$ is normalized to a range of [0, 1]. The radiation amplitude of each radiation element required for generating a wave propagating in the direction $(\phi_0,\varphi_0)$ can then be mathematically formulated as:"}
\begin{equation}
	m_{x,y}(r^q_{x,y},\phi_0,\varphi_0) = \frac{\mathrm{Re}[\Psi_{intf}(\mbf{r}_{x,y}^{q},\phi_0,\varphi_0)]+1}{2}.
	\label{Eq:holo_4}
\end{equation}
{According to~\eqref{Eq:holo_4}, it is observed that the particular elements whose reference waves closely match the object wave (i.e., have a large amplitude) are tuned to emit strongly. Conversely, the radiation elements that do not closely match are tuned down or even turned off. For the multi-beam design of the holographic pattern, the amplitude of each element is then determined accordingly as}
\begin{equation}
	m_{x,y}(r^q_{x,y},\phi_0,\varphi_0) =  \sum_{k=1}^{K} \sum_{q=1}^{N_{RF}} a_{k,q}  m_{x,y}(r^q_{x,y},\phi_k,\varphi_k) \;\;\forall(x,y).
	\label{Eq:holo_5}
\end{equation}
{Here, $a_{k,q}$ denotes the amplitude ratio for the beam directed towards the $k$-th UE from the $q$-th feed, and it must satisfy the condition:  $ \sum_{k=1}^{K} \sum_{q=1}^{N_{RF}} a_{k,q} = 1$. This condition ensures that the total amplitude contribution across all directions and feeds sums to 1, which also ensures $0 \leq  m_{x,y} \leq 1$.}

\subsection{Received Signal at the k-th UE}

\noindent
The intended signal vector for $K$ UEs is $\mbf{s} \in \mathbb{C}^{K \times 1}$. Consequently, the signals {transmitted} from the BS are expressed {by the vector} $\mbf{Fs}$. Therefore, the signal received by the $k$-th UE, denoted as ${y}_k$, can be expressed as follows:
\begin{align} 
	{y}_{k}&=\mbf {h}_{tot,k}^H\mbf {M}_v\mbf {F}_{}\mbf {s}_{}+ {n}_{k}, \label{Eq:received_signal} \\		
	&=\underbrace {\mbf {h}_{tot,k}^H\mbf {M}_v\mbf {f}_{k} {s}_{k}}_{\text{Intended signal}}+\underbrace{\mbf {h}_{tot,k}^H\mbf {M}_v\sum _{k'\neq k}\mbf {f}_{k'} {s}_{k'}}_{_{\text{Interference}}}+ \underbrace{{n}_{k}}_{_{\text{Receiver noise}}}, 
	\label{Eq:received_signal2}
\end{align}
where $n_k \thicksim  \mathcal{CN}(0,\sigma^2)$ denotes the additive white Gaussian noise (AWGN) at the $k$-th UE.

\subsection{Mutual Coupling}

{Mutual coupling in wireless communication systems refers to the interaction between closely placed antennas in an array, where the electromagnetic fields generated by one antenna affect the performance of neighboring antennas~\cite{Tareq_2024_IEEEWCL_Mutualcoup,Debbah_2024_IEEEJSAC_holo_Mutual}. Let the effects of mutual coupling among RHS elements be fully captured by the coupling matrix $\mbf{C}$, which generally depends on the positions and radiation power patterns of the antenna elements. If the mutual coupling effect is not considered, the coupling matrix $\mbf{C}$ is assumed to be an identity matrix ($\mbf{I}_{N_t \times N_t}$).
In the RHS transmitter model associated with ${N_t}$ discrete elements, the  electromagnetic response including mutual coupling is given by
\begin{equation}
	{\mbf{M}_v} = \mbf{C}  {\mbf{M}}{{\mathbf{V}}}, \;\;     
	\begin{cases}
	     \mbf{C}= \mbf{I} \;\; \textrm{Without mutual coupling}\\
		\mbf{C}\neq \mbf{I} \;\; \textrm{With mutual coupling}. \end{cases}
	\label{Eq:MC_holo_matrix}
\end{equation}	
With mutual coupling, the signal received by the $k$-th UE is given by
	\begin{align} 
		{y}_{k}&=\mbf {h}_{tot,k}^H \mbf{C} \mbf {MV}\mbf {F}_{}\mbf {s}_{}+ {n}_{k}.  
		\label{Eq:MC_received_signal}
	\end{align}
We consider the $N_t$ RHS elements at the BS to be isotropic, with equal spacing between the successive elements~\cite{Marzetta_2022_IEEETWC_Holofourier}. The elements in the mutual coupling matrix $\mbf{C}$ are modeled according to~\cite{Branislav_2023_IEEETWC_holo_beamforming},
and the coupling matrix is given by
\begin{equation}
	\mbf{C} = \mbf{D}^{-1/2},
	\label{Eq:MC_matrix}
\end{equation}
where the matrix $\mbf{D} =  \{c_{n',n}\}$ is an $N_t \times N_t$ matrix, and for the element positions $\{\mbf{t}_n \;|\; n = 1, 2,\hdots, N_t\}$, $c_{n',n}$ is given by  
\begin{equation}
{	c_{n',n} = \textrm{sinc}\;(2\|\mbf{t}_{n'}-\mbf{t}_{n}\|_2/\lambda),}
\end{equation}
where $\|\mbf{t}_{n'}-\mbf{t}_{n}\|_2$ represents the distance between the RHS elements $n'$ and $n$.  It can be observed that for an antenna array having a given geometrical deployment, where all the RHS antenna elements are isotropic, the coupling matrix $\mbf{C}$ in~\eqref{Eq:MC_matrix} is deterministic and
only has to be computed once, offline.
This approach is based on coupling-agnostic transceiver designs, which may result in a loss of performance. The corresponding simulations, with and without the effects of mutual coupling, are presented in the simulation results section for the proposed and existing methods. For simplicity, the coupling matrix $\mbf{C}$ is assumed to be the identity matrix $\mbf{I}$ for the rest of the paper, as this does not affect the analysis.}

\subsection{Beamformer Design}

\noindent
The achievable rate of the $k$-th UE is given by
\begin{equation}
	R_{k}=\log _{2}\left ({1+\frac {|\mbf {h}_{tot,k}^H\mbf {M}_v\mbf {f}_{k}|^{2}}{\sigma ^{2}+\sum _{k'\neq k}|\mbf {h}_{tot,k}^H\mbf {M}_v\mbf {f}_{k'}|^{2}}}\right).
	\label{Eq:achieve_rate}
\end{equation}

\noindent
Our objective is to maximize the achievable {DL sum-rate} through the optimization of the digital beamformer matrix $\mbf{F}$, the holographic beamformer matrix $\mbf{M}$, and the RIS phase shift matrix $\bols{\Theta}_{RIS}$. The optimization problem is formulated as follows:
\begin{align}
	&\max _{\{\mbf{F},\mbf{M},\bols{\Theta}_{RIS}\}}\sum _{k=1}^{K}R_{k}\nonumber\\&\quad \, s.t.\;\; \mrm{Tr}(\mbf {M}\mbf{VF}\mbf {F}^{H}\mbf {V}^{H}\mbf {M}^{H})\leq P_{T}, \nonumber\\&\hphantom {\quad \, s.t. } \;\; 0\leq  {m}_{x,y}\leq 1,\quad \forall x,y,  \nonumber
	\\&\qquad\quad |\bols{\Theta} _{RIS}(i,i)|^{2} = 1,\quad \forall  i
	\label{Eq:sum-rate_opti}
\end{align}
where $P_T$ represents the total transmit power {available} at the BS. The amplitudes ${m}_{x,y}$ of the holographic beamformer are constrained to a maximum value of 1. Additionally, the constraint on $\bols{\Theta} _{RIS}$ specifies that {the} RIS elements reflect the impinging signal without any loss of energy. The problem described in~\eqref{Eq:sum-rate_opti} is non-convex due to the product of optimization variables. Therefore, we rely on the AM algorithm, where in each step, we solve a subproblem, which is convex.


\section{Alternating Maximization Algorithm}

In this section, we implement an AM algorithm~\cite{Bin_2021_IEEETVT_AMalgo} to address the sum-rate maximization problem {of} the RHS-RIS-aided multi-UE communication system. The {considered} joint optimization problem is tackled by decomposing it into three subproblems: the digital beamforming subproblem, the holographic beamforming subproblem, and the RIS phase shift matrix optimization subproblem. In the AM approach, the fundamental concept involves decomposing the optimization variables into multiple blocks. Subsequently, each block is updated following specific rules, while keeping the remaining blocks fixed at their {previous} updated values.

\subsection{Problem Decomposition}

To address the optimization problem described in ~\eqref{Eq:sum-rate_opti} and achieve the best sum-rate, we decouple the nonconvex optimization problem into three subproblems outlined below. We solve these subproblems sequentially to obtain the optimal values for  $\mbf {F}$, $\mbf {M}$  and $\bols{\Theta}_{RIS}$.

\subsubsection{Digital Beamforming}
Given the holographic beamformer matrix $\mbf{M}$ and the RIS phase shift matrix $\bols{\Theta}_{RIS}$, the digital beamforming subproblem can be {formulated} as
\begin{equation} 
	{\mathcal{P}_1}: \;\; \max _{\{\mathbf {F}\}}\sum _{k=1}^{K}R_{k},\quad s.t. \;
	\mrm{Tr}(\mbf {M}_v\mbf{F}\mbf {F}^{H}\mbf {M}_v^{H})\leq P_{T}.
\end{equation}

\subsubsection{Holographic Beamforming}
Given the optimal digital beamformer {matrix} $\mbf{F}$ of ${\mathcal{P}_1}$ and the RIS phase shift matrix $\bols{\Theta}_{RIS}$, the subproblem can be {formulated} as
\begin{equation} 
	{\mathcal{P}_2}: \;\; \max _{\{\mathbf {M}_{}\}}\sum _{k=1}^{K}R_{k},\quad s.t. \; 0\leq  {m}_{x,y}\leq 1, \quad \forall x,y.
\end{equation}

\subsubsection{RIS phase optimization}
Given the optimal digital beamformer {matrix} $\mbf{F}$ of ${\mathcal{P}_1}$ and the optimal holographic beamformer {matrix} $\mbf{M}$ of ${\mathcal{P}_2}$, the RIS phase {shift} matrix design subproblem can be {formulated} as
\begin{equation}
	{\mathcal{P}_3}: \;\; \max _{\{\bols{\Theta}_{RIS}\}}\sum _{k=1}^{K}R_{k},\quad s.t. \; |\bols{\Theta} _{RIS}(i,i)|^{2} = 1,\quad \forall  i.
\end{equation}

The AM algorithm iterates between ${\mathcal{P}_1}$, ${\mathcal{P}_2}$, and ${\mathcal{P}_3}$ to obtain the optimal values for $\mbf {F}$, $\mbf {M}$, and $\bols{\Theta}_{RIS}$.

\subsection{Digital {Beamformer} Design ($\mathcal{P}_1$)}
\noindent The digital beamforming subproblem, obtained by substituting $R_k$ from ~\eqref{Eq:achieve_rate} into ${\mathcal{P}_1}$ is  presented as follows:
\begin{align} 
	{\mathcal{P}_1}: \;\;  \max \limits_{\{\mbf {F}_{}\}}&\sum _{k=1}^{K}\mrm{log}\left ({1+\frac {|\mbf {h}_{tot,k}^H\mbf {MV}\mbf {f}_{k}|^{2}}{\sigma ^{2}+\sum _{k'\neq k}|\mbf {h}_{tot,k}^H\mbf {MV}\mbf {f}_{k'}|^{2}}}\right),\nonumber \\ 
	\quad &s.t. \; 	\mrm{Tr}(\mbf {MV}\mbf{F}\mbf {F}^{H}\mbf {MV}^{H})\leq P_{T}. 
\end{align}
The problem $\mathcal{P}_1$ is a well-known {one}, typically addressed using zero-forcing beamforming. In~\cite{Shlomo_2008_IEEETSP_ZFBF}, it has been demonstrated that zero-forcing beamforming can achieve near-optimal solutions {at} low complexity. Therefore, we opt {for employing} zero-forcing beamforming along with power allocation as the low-dimensional digital beamformer at the BS. This decision is motivated by the need to mitigate {the} inter-user interference. The digital beamformer {matrix} can be expressed as follows:
\begin{equation} 
	\mbf {F}=\mbf {Q}^{H}(\mbf {QQ}^{H})^{-1}\mbf {P}^{\frac {1}{2}}=\widetilde {\mbf {F}}\mbf {P}^{\frac {1}{2}},
	\label{Eq:DBF_ZF}
\end{equation}
\noindent where $\mbf{Q}=[\mbf{V}^H\mbf{M}^H\mbf{h}_{tot,1},\hdots,\mbf{V}^H\mbf{M}^H\mbf{h}_{tot,K}] \in \mbb{C}^{K\times N_{RF}}$, $\mbf{P}=\mrm{diag}(p_1,\hdots,p_K)$ is a diagonal matrix, and $p_k$ represents the power {received} at the $k$-th UE. Using~\eqref{Eq:DBF_ZF} and by leveraging the properties of zero-forcing beamforming, i.e., $\mbf{h}_{tot,k}^H\mbf{MV}\mbf{f}_{k}=\sqrt{p_k}$ and $\mbf{h}_{tot,k}^H\mbf{MV}\mbf{f}_{k'}=0, \forall k'\neq k$, the digital beamforming subproblem {may be simplified} into a power allocation problem, as presented below:
\begin{align}
	&\max _{\{p_{k}\}} \sum _{k=1}^{K}\log _{2}\left ({1+\frac {p_{k}}{\sigma ^{2}}}\right) \nonumber,\\&s.t.\; {\mrm {Tr}}\left({\mathbf {P}^{\frac {1}{2}}\widetilde {\mbf {F}}^{H}\mbf {V}^{H}\mbf {M}^{H}\mbf {MV}\widetilde {\mbf {F}}\mbf {P}^{\frac {1}{2}}}\right)\leq P_{T}, \quad p_{k}\geq 0.
\end{align}
The optimal {$p_k^*$} can be obtained by {the} water-filling algorithm~\cite{Pramod_2005_textbook_fundamentals} as
\begin{equation} 
	p_{k}^{\ast }=\frac {1}{\mu _{k}}\max \left\{{\frac {1}{\epsilon}-\mu _{k}\sigma ^{2},0}\right\},
	\label{Eq:Power_ZF}
\end{equation}
where $\mu_k$ is the $k$-th diagonal element of $\widetilde {\mbf {F}}^{H}\mbf {V}^{H}\mbf {M}^{H}\mbf {MV}\widetilde {\mbf {F}}$, and $\epsilon$ is a {normalization} factor satisfying $\sum _{k=1}^{K}$max$\{{\frac{1}{\epsilon}}-\mu_k\sigma^2,0\}=P_T$. 
By obtaining {the} {$p_k^*$} {values} from~\eqref{Eq:Power_ZF}, we obtain the matrix $\mbf{P}$, {upon} substituting the matrix $\mbf{P}$ into $\mbf {F}=\widetilde {\mbf {F}}\mbf {P}^{\frac {1}{2}}$, we can then derive the optimal digital beamformer matrix $\mbf{F}$.

\subsection{Holographic {Beamformer} Design ($\mathcal{P}_2$)}

The optimization problem described in~\eqref{Eq:sum-rate_opti} differs from {the} traditional phase-controlled analog beamforming design. The objective function in $\mathcal{P}_2$ is nonconvex as it involves optimization variables in fractional form. Consequently, {the} existing algorithms like {the} semi-definite programming (SDP)\cite{Lajos_2022_IEEETVT_RIS_ND} and {the} gradient ascent algorithm\cite{Wei_2016_IEEEJSTSP_hybrid_GradAsc} are unable to handle this. To address this challenge, we employ fractional programming~\cite{Wei_2018_IEEETSP_fractionalprogram} {based} optimization, which solves the problem $\mathcal{P}_2$ as a series of optimization problems discussed in this section:
 \begin{align} 
	{\mathcal{P}_2}: \;\;  \max \limits_{\{\mbf {M}_{}\}}&\sum _{k=1}^{K}{\log}\left ({1+\frac {|\mbf {h}_{tot,k}^H\mbf {MV}\mbf {f}_{k}|^{2}}{\sigma ^{2}+\sum _{k'\neq k}|\mbf {h}_{tot,k}^H\mbf {MV}\mbf {f}_{k'}|^{2}}}\right), \nonumber \\ 
	\quad &s.t. \; 0\leq  {m}_{x,y}\leq 1, \quad \forall x,y, 
\end{align}
\begin{align} 
	{\mathcal{P}_{2.1}}: \;\; \max \limits_{\{\mbf {M}_{}\}}&\sum _{k=1}^{K}\mrm{}\left ({\frac {|\mbf {h}_{tot,k}^H\mbf {MV}\mbf {f}_{k}|^{2}}{\sigma ^{2}+\sum _{k'\neq k}|\mbf {h}_{tot,k}^H\mbf {MV}\mbf {f}_{k'}|^{2}}}\right),\nonumber\\ 
		\quad &s.t. \; 0\leq  {m}_{x,y}\leq 1, \quad \forall x,y. 
\end{align}
The subproblem ${\mathcal{P}_2}$ is not convex. 
To address ${\mathcal{P}_2}$, we approximate the logarithmic term ${\log}(1+{x})$ {by} its first-order Taylor series expansion, which gives ${x}$. Therefore, ${\mathcal{P}_{2.1}}$ is solved to tackle ${\mathcal{P}_2}$.

\noindent {Upon} reformulating $\mathcal{P}_{2.1}$ as {follows:}
\begin{align} 
	{\mathcal{P}_{2.1}}: \;\; \max \limits_{{\bols{0} \preccurlyeq {\mathbf{m}} \preccurlyeq \bols{1}}}\sum \limits_{k=1}^{K}\left({\frac{{{{\mbf{m}}^T}\Re \left( {{\mbf{\Sigma }}_k^{{{\mathcal{P}}_2}}} \right){\mathbf{m}}}}{{{{\mathbf{m}}^T}\Re \left( {\widetilde {\mathbf{\Sigma }}_k^{{{\mathcal{P}}_2}}} \right){\mathbf{m}} + \sigma ^2}}}\right) ,
 \end{align}
\noindent where {we have} $\mbf{m} = \mbf{M}^T\mbf{1}_{N_t}$, the matrices ${{\mbf{\Sigma }}_k^{{{\mathcal{P}}_2}}}$ and ${\widetilde {\mathbf{\Sigma }}_k^{{{\mathcal{P}}_2}}}$ are defined as
\begin{align}
	&\mbf{\Sigma }_k^{{\mathcal{P}}_2}= \mrm{diag}(\mbf{Vf}_k)\mbf{h}_{tot,k}\mbf{h}_{tot,k}^H\mrm{diag}(\mbf{Vf}_k)^H,\\
	&{\mbf{\widetilde\Sigma }_k^{{\mathcal{P}}_2}}=
	\mrm{diag}(\mbf{Vf}_{k'})\mbf{h}_{tot,k}\mbf{h}_{tot,k}^H\mrm{diag}(\mbf{Vf}_{k'})^H.
\end{align}
Note that $\mathcal{P}_{2.1}$ is {still} nonconvex due to the fractional quadratic objective function. Hence, we simplify it {by harnessing} the {first-order Taylor} approximation as
\begin{align} 
	&{\mathcal{P}_{2.2}}: \;\; \nonumber \\ 
	&\max \limits_{{\bols{0} \preccurlyeq {\mathbf{m}} \preccurlyeq \bols{1}}}&\sum \limits_{k=1}^{K}\bigg({\frac{{2{{\mathbf{m}}^{{{(0)}^T}}}\Re \left( {{\mathbf{\Sigma }}_k^{{{\mathcal{P}}_2}}} \right){\mathbf{m}} - {{\mathbf{m}}^{{{(0)}^T}}}\Re \left( {{\mathbf{\Sigma }}_k^{{{\mathcal{P}}_2}}} \right){{\mathbf{m}}^{(0)}}}}{{{{\mathbf{m}}^T}\Re \left( {\widetilde {\mathbf{\Sigma }}_k^{{{\mathcal{P}}_2}}} \right){\mathbf{m}} + \sigma ^2}}}\bigg),\nonumber \\ 
	\label{Eq:RHS_Solution}
\end{align}
where $\mathcal{P}_{2.2}$ is a standard fractional maximization problem that can be solved using {the popular} Dinkelbach-based method~\cite{Bhavani_2022_IEEEISJC_RIS_FP}.
{The Algorithm~\ref{Algo:Dinkelbach} in Appendix  summarizes the solving method for problem $\mathcal{P}_{2.2}$.
}


\begin{algorithm}[t] 
	\DontPrintSemicolon
	\SetAlgoLined
	\SetKwInput{kwInit}{Init}
	\SetKw{Kw}{Variables}
	\KwIn{Initialize digital beamformer matrix $\mbf{F}^{(0)}$, holographic beamformer matrix $\mbf{M}^{(0)}$ and RIS phase shift matrix $\bols{\Theta}_{RIS}^{(0)}$. Set $t$ = 1. }  
	
	\KwOut{Optimal $\mbf{F}$, $\mbf{M}$ and $\bols{\Theta}_{RIS}$}
	
	\hspace{1.5mm}Compute the digital beamformer matrix $\mbf{F}^{(t)}$ using $\mbf{M}^{(t-1)}$, $\bols{\Theta}_{RIS}^{(t-1)}$ and~\eqref{Eq:DBF_ZF}.  \;
	
	\hspace{1.5mm}Compute the holographic beamformer matrix $\mbf{M}^{(t)}$ using $\mbf{F}^{(t)}$, $\bols{\Theta}_{RIS}^{(t-1)}$ and~\eqref{Eq:RHS_Solution}. \;
	
	\hspace{1.5mm} Compute the RIS phase shift matirx $\bols{\Theta}_{RIS}^{(t)}$ using $\mbf{F}^{(t)}$, $\mbf{M}^{(t)}$ and the Riemannian conjugate gradient algorithm to solve~\eqref{Eq:RIS_Formulation}.\;
	
	Set $t = t + 1$; \;
	
	\tbf{repeat steps 1 to 4}
	
	\tbf{until} The value of the objective function in~\eqref{Eq:sum-rate_opti} converges {or maximum iteration reached.}
	
	\caption{{Alternating maximization algorithm}}
	\label{Algo:AM_algorithm}
\end{algorithm} 

\begin{figure*}[hbt!]
	\centering
	\includegraphics[width=6.7in]{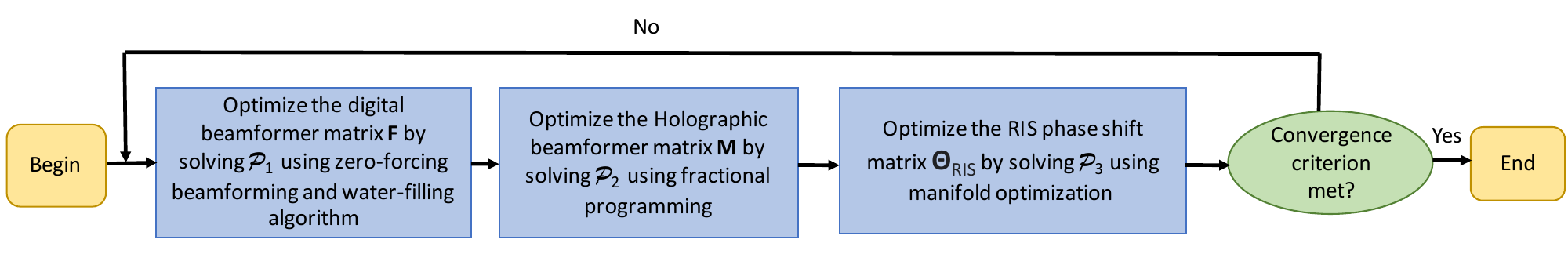}
	\caption{\small Flow chart of the proposed AM algorithm for sum-rate.}
	\label{Fig:Flow_chart_AM}
\end{figure*}

\vspace{-1.5mm}

\subsection{RIS Phase {Shift} Matrix Design ($\mathcal{P}_3$)}
Next we focus on the RIS phase optimization subproblem. For ease of representation, we define the effective channels for the direct link and the RIS link. From~\eqref{Eq:received_signal} and~\eqref{Eq:received_signal2} we have
\begin{align} 
	{y}_{k}&=\mbf {h}_{tot,k}^H\mbf {M}_v\mbf {F}_{}\mbf {s}_{}+ {n}_{k}, \nonumber \\
	&=\underbrace {\mbf {h}_{d,k}^H\mbf {M}_v\mbf {F}_{}\mbf {s}_{}}_{\text {Direct link}}+ \underbrace { \mbf {h}_{R,k}^H\bols{\Theta}_{RIS} \mbf{G}_R\mbf {M}_v\mbf {F}_{}\mbf {s}_{} }_{{\text {RIS-}}{\text {aided}}~{\text {link}}} +{n}_{k}, \nonumber\\
	&=(\mbf {h}_{d,k}^H+ \mbf {h}_{R,k}^H\bols{\Theta}_{RIS} \mbf{G}_R)\mbf {M}_v\mbf {F}_{}\mbf {s}_{}+{n}_{k},\nonumber\\
	&=(\mbf {h}_{d,k}^H+ \mbf {h}_{R,k}^H\bols{\Theta}_{RIS} \mbf{G}_R)\mbf {M}_v  \sum _{k=1}^{K} \mbf {f}_{k} {s}_{k}+{n}_{k}.
	\label{Eq:RIS_expand}	
\end{align}
To make expression~\eqref{Eq:RIS_expand} more tractable, we further define $\bols{\theta}_{RIS}$ = $[e^{j\theta_{1}}, \hdots,e^{j\theta_{N_{RIS}}}]^H$ and $\mbf{H} _{R,k} = \mrm{diag}(\mbf {h}_{R,k}^H) \mbf{G}_R$.
The received signal $y_k$ is equivalently represented as
\begin{equation} 
	{y}_{k}= \left ({\mbf {h}_{{d,k}}^{H}+{\bols{\theta}}_{RIS}^{H} \mbf {H}_{{R,k}}}\right)\mbf{M}_v \sum _{k=1}^{K} \mbf {f}_{k} s_{k} +n_{k}.
\end{equation}
For the sake of {notational simplicity}, let {us} define the effective channels for the direct link and the RIS link as follows:
\begin{align} 
	\mbf{a}_{k',k}= \;&\mbf{H}_{R,k} \mbf{M}_v \mbf {f}_{k'},\\ 
	b_{k',k}= \;&\mbf{h}_{d,k}^{H} \mbf{M}_v \mbf {f}_{k'},
\end{align}
respectively. The phase optimization subproblem ${\mathcal{P}_3}$ can be expressed as a function of $\bols\theta_{RIS}$ {as follows:}
\begin{align}
	 \mathcal{P}_3 : \;\; &\max \limits_{\{\bols \theta_{RIS}\} } \;  f_{{}}({\boldsymbol \theta_{RIS} }) \nonumber\\
	 & {s.t.}~|\bols{\theta} _{RIS}(i)|^{2} = 1, \quad \forall i=1,\cdots,N_{RIS}
	 \label{Eq:RIS_Formulation}
\end{align}
where {we have:}
\begin{align}  
	f_{\mathrm{}}({\boldsymbol \theta_{RIS} })=&\sum _{k=1}^{K} R_k, \nonumber\\
	=&\sum _{k=1}^{K} \log_2 \left({1+\frac {\left |{{\boldsymbol \theta^H_{RIS} } \mbf {a}_{k,k}+b_{k,k} }\right |^{2} } {\sum \limits_{k'\neq k} \left |{{\boldsymbol \theta^H_{RIS} } \mbf {a}_{k',k}+b_{k',k} }\right |^{2}+\sigma _{}^{2}}}\right).
\end{align}
It can be observed that $f(\bols{\theta}_{RIS})$ is both continuous and differentiable. Additionally, the constraint set of $\bols{\theta}_{RIS}$ forms a complex circle manifold. Consequently, the stationary solution of $\mathcal{P}_3$ can be obtained via the Riemannian conjugate gradient algorithm~\cite{Rodolphe_2014_JournML_manopt}. {The Riemannian conjugate gradient algorithm exhibits better robustness to initialization than the family of gradient-based methods in Euclidean spaces. Due to the constant modulus constraint in~\eqref{Eq:RIS_Formulation}, the problem is known to be NP-hard. Other approaches, such as the semi-definite relaxation with randomization, and the majorization-minimization framework,
could also be used, but they incur relatively high computational costs due to the non-convex constant modulus constraint. The Riemannian conjugate gradient algorithm naturally incorporates the physical constraints of RIS systems, such as the constant modulus constraint, making it particularly effective in handling the complex non-convex optimization problems associated with RIS phase shifts on a manifold.} Conceptually, the Riemannian conjugate gradient algorithm involves three key steps in each iteration:

\subsubsection{Compute {the} Riemannian Gradient}
For a smooth function $f$ defined on a Riemannian manifold $\mathcal{M}$, the Riemannian gradient (${\mathrm{grad}}f$) is the orthogonal projection of the Euclidean gradient $\nabla f$ onto the complex circle {formulated as}:
\begin{equation} 
	{\mathrm{ grad}} f_{\mathrm{}}= \nabla f_{\mathrm{}}-{\mathrm{ Re}} \left \{{\nabla f_{\mathrm{}} \circ {\boldsymbol \theta^\ast_{RIS} } }\right \}\circ {\boldsymbol \theta_{RIS} },
\end{equation}
where the Euclidean gradient is
\begin{equation} 
	\nabla f_{\mathrm{}}=\sum _{k=1}^{K} 2 \mbf {A}_{k},
\end{equation}
{in conjunction} with 
\begin{align}
	&\hspace{-0.5pc}\mbf {A}_{k}=\frac {\sum _{k'} \mbf {a}_{k',k} \mbf {a}_{k',k}^{{H}}{\boldsymbol \theta_{RIS} }+ \sum _{k'} \mbf {a}_{k',k} b_{k',k}^\ast } {\sum _{k'} \left |{{\boldsymbol \theta^H_{RIS} } \mbf {a}_{k',k}+b_{k',k} }\right |^{2}+\sigma _{}^{2}}\nonumber\\&\qquad \qquad \displaystyle {\quad -\frac {\sum _{k' \neq k} \mbf {a}_{k',k} \mbf {a}_{k',k}^{{H}}{\boldsymbol \theta_{RIS} }+ \sum _{k' \neq k} \mbf {a}_{k',k} b_{k',k}^\ast } {\sum _{k' \neq k} \left |{{\boldsymbol \theta^H_{RIS} } \mbf {a}_{k',k}+b_{k',k} }\right |^{2}+\sigma _{}^{2}}.} 
\end{align}

\subsubsection{Search Direction}
In general, the search direction is given by the {opposite} of the Riemannian gradient. The tangent vector conjugate {of} $\mathrm{grad}f$ gives the search direction:
\begin{equation}
	\bols{\eta}=-{\mathrm{ grad}} f_{\mathrm{}}+\beta _{1} {\mathcal T}({\bar {\bols {\eta}}}),
\end{equation}
where $\Tau(.)$ is the vector transport function defined as
\begin{equation} 
	{\mathcal T}(\bols {\eta})={\bar {\bols {\eta}}}- {\mathrm{ Re}} \left \{{\bols {\eta}\circ {\boldsymbol \theta ^\ast_{RIS}} }\right \}\circ {\bols \theta_{RIS} },
\end{equation}
{while} $\beta_{1}$ is the conjugate gradient update parameter, and ${\bar{\bols{\eta}}}$ is the previous search direction.
\subsubsection{Retraction}
 The retraction step updates the current point on the manifold using the search direction to ensure that the next {iteration} remains on the manifold. By projecting the tangent vector back to the complex circle manifold we get
\begin{equation}
	{\boldsymbol \theta }_{RIS}(n) \leftarrow \frac {({\boldsymbol \theta_{RIS} }+\beta _{2} \bols {\eta})_{n}} {|({\boldsymbol \theta_{RIS} }+\beta _{2} \bols {\eta})_{n}|},
\end{equation}
where ${\boldsymbol \theta }_{RIS}(n)$ represents the $n$-th component of the vector ${\boldsymbol \theta }_{RIS}$ and $\beta_{2}$ is the Armijo step size, {which determines the size of the step taken along the search direction in each iteration.}

To gain a clear understanding of the variable updating process in each step, {please} refer to the detailed {procedure formulated} by {Algorithm} in~\ref{Algo:AM_algorithm}.

\subsection{{Computational Complexity}}
{In this section we first calculate the complexity of the algorithms for three subproblems separately and then compute the overall computational complexity of the alternating maximization algorithm in~\ref{Algo:AM_algorithm}. 
\begin{itemize}
	\item For the update of the digital beamforming matrix $\mbf{F}$, we use the zero-forcing algorithm to mitigate interference by applying a pseudo-inverse of the channel matrix to the received signal. The complexity of this step is $\mathcal{O}(K^2 N_{RF})$. The normalization factor $\epsilon$ is obtained by solving the equation $\sum _{k=1}^{K}$max$\{{\frac{1}{\epsilon}}-\mu_k\sigma^2,0\}=P_T$, which has a complexity of $\mathcal{O}(K^2)$. Following that, the optimal received power
	$p_k^*$ is obtained by~\eqref{Eq:Power_ZF} with a complexity of $\mathcal{O}(K)$.
	Therefore, the total complexity of the digital beamforming is $\mathcal{O}(K^2N_{RF})$. 
	\item At each inner iteration of the Dinkelbach-based method, the problem is solved by CVX using the interior-point method, which generally has a complexity of $\mathcal{O}(N_t^3)$ for the update of the holographic beamformer matrix $\mbf{M}$. Given a total of $l_M$ inner iterations in  Algorithm~\ref{Algo:Dinkelbach}, the complexity is $\mathcal{O}(l_MN_t^3)$. 
	\item To update the RIS phase shift matrix $\bols{\Theta}_{RIS}$, we use the Riemannian conjugate gradient algorithm. The complexity of the Riemannian conjugate gradient algorithm is dominated by the computation of the Euclidean gradient, which is $\mathcal{O}(K^2N_{RIS}^2)$. The retraction step also requires iteratively searching for ${\beta}_2$ with a complexity of $\mathcal{O}(K^2N_{RIS})$, which can be ignored when $N_{RIS}$ is large.
\end{itemize}
Therefore the total complexity for each outer iteration of the proposed AM algorithm is $\mathcal{O}(K^2N_{RF}+l_MN_t^3+K^2N_{RIS}^2)$. }




\subsection{{Convergence}}
{In the AM Algorithm~\ref{Algo:AM_algorithm} convinced for  digital beamforming, the sum-rate ($R_{sum}$) in~\eqref{Eq:achieve_rate}
becomes non-decreasing after optimizing the digital beamformer matrix $\mbf{F}^{}$, given the holographic beamformer $\mbf{M}^{(t-1)}$ and RIS phase shift matrix $\bols{\Theta}_{RIS}^{(t-1)}$ in the $t$-th iteration, i.e.,  
\begin{equation}
	R_{sum}(\mbf{F}^{(t)}, \mbf{M}^{(t-1)},\bols{\Theta}_{RIS}^{(t-1)}) \geq R_{sum}(\mbf{F}^{(t-1)}, \mbf{M}^{(t-1)},\bols{\Theta}_{RIS}^{(t-1)}).
\end{equation}
Secondly, given the digital beamformer matrix $\mbf{F}^{(t)}$ and the RIS phase shift matrix $\bols{\Theta}_{RIS}^{(t-1)}$, the holographic beamformer matrix $\mbf{M}^{(t-1)}$ is optimized using an algorithm proposed in~\ref{Algo:Dinkelbach}, so that we obtain
\begin{equation}
	R_{sum}(\mbf{F}^{(t)}, \mbf{M}^{(t)},\bols{\Theta}_{RIS}^{(t-1)}) \geq R_{sum}(\mbf{F}^{(t)}, \mbf{M}^{(t-1)},\bols{\Theta}_{RIS}^{(t-1)}).
\end{equation}
After updating the two variables, given the digital beamformer matrix $\mbf{F}^{(t)}$ and the holographic beamformer matrix $\mbf{M}^{(t)}$, the RIS phase shift matrix $\bols{\Theta}_{RIS}^{(t-1)}$ is optimized using manifold optimization. Consequently, we obtain
\begin{equation}
	R_{sum}(\mbf{F}^{(t)}, \mbf{M}^{(t)},\bols{\Theta}_{RIS}^{(t)}) \geq R_{sum}(\mbf{F}^{(t)}, \mbf{M}^{(t)},\bols{\Theta}_{RIS}^{(t-1)}).
\end{equation}
This means that in each update step of the proposed algorithm, the objective function value (i.e., sum-rate) does not decrease. Additionally, the sequence of objective function values obtained during the iteration steps is both monotonic and bounded, ensuring that the overall algorithm will converge.}

\subsection{{Power Consumption and Hardware Cost}}
{In this section, we compare the power consumption and hardware cost of phased array and RHS systems. Let us consider the hardware cost of a phased array module as $\xi_{ph}$ and the hardware cost of an RHS module as $\xi_{rhs}$. We define the cost ratio  $C_r$ as the ratio of the phased array's cost to the RHS cost, i.e., $C_r$ = $\frac{\xi_{ph}}{\xi_{rhs}}$. For a phased array system with $N_t$ antennas, the hardware cost is $N_t C_r \xi_{rhs}$. Similarly, for an RHS system having $N_t$ elements, the total hardware cost is $N_t \xi_{rhs}$.  
In the RHS prototype~\cite{Commware_2019_holographic_power}, it is shown that in array grids, RHSs typically have about $2.5$ times as many elements as a phased array system having the same antenna directivity. 
Additionally, the radiation power to total power consumption for phased array and RHS systems is $4$\% and $25$\%, respectively. In this case, the total hardware cost of an RHS system is $2.5 N_t \xi_{rhs}$ and for the phased array system, it is  $N_t C_r \xi_{rhs}$. Given that the typical value of $C_r$ is $10$~\cite{Commware_2019_holographic_power}, the hardware cost of the phased array system is higher than that of the RHS system. This demonstrates that RHS provides a powerful solution to reduce hardware costs with lower power consumption, while guaranteeing better directivity in practice.}

\section{Simulation Results}
In this {section}, we present simulation results {for characterizing} the performance of the proposed approach. {We  focus exclusively on the far-field region, as both the RIS and UEs are positioned beyond the Rayleigh distance $(D_R = 2D^2/\lambda)$ from the BS,
where $D$ denotes the array aperture.} The AM algorithm put forward in this study leverages perfect channel state information (CSI) to achieve the optimal sum-rate for the system.
\begin{table}[hbt!]
	\caption{ {Simulation Parameters}}
	\centering
	\begin{tabular}{|l |l |}
		\hline
		$\mbf{Parameters}$  & $\mbf{Values}$  \\ \hline
		{Number of RHS elements} ($N_t$) &  16 to 64    \\ \hline		
		Number of RIS elements ($N_{RIS}$) & 20 to 100   \\ \hline
		Number of RF chains ($N_{RF}$)  & 8  \\ \hline
		Number of UEs ($K$) & 2 to 4     \\ \hline
		Number of antennas at each UE  & 1     \\ \hline
		Number of multipaths  & 10     \\ 
		 ($L_d, L_{br}, L_{ru}$) &      \\ \hline
		Operating frequency (${f}_c$) in GHz  & 28     \\ \hline
		Distance between the RIS &   \\ 
		elements ($d = \lambda/2$) in cm & 0.53 \\ \hline
		Distance between the RHS &   \\ 
		elements ($d = \lambda/4$) in cm & 0.265 \\ \hline
		Transmit power of the BS   &  3 to 30    \\ 
		(${P}_T$) in W  &    \\ \hline
		{Noise power} (dBm) & -90 \\ \hline
		{Path loss for LOS link (dB) } &  \hspace{-3mm} $61.4+20\;\mrm{log}(d_i)+5.8 $      \\ \hline
		{Path loss for  NLOS link (dB) } &  \hspace{-1.7mm}$72+29.2\;\mrm{log}(d_i)+8.7 $      \\ \hline
	\end{tabular}
	\label{Tab:Simulation Parameters}
	\end{table}

%
%
%
%
%

\begin{figure}[hbt!]
	\centering
	\includegraphics[width=0.9\linewidth]{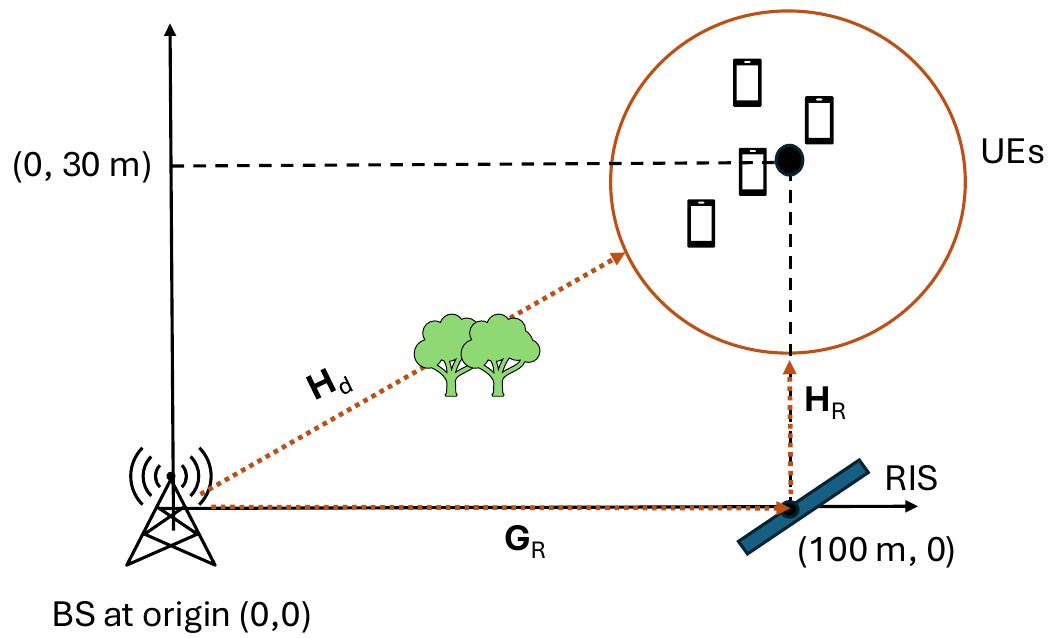}
	\caption{\small Overview of the simulated RHS-RIS system having $K$ UEs with one receive antenna, a BS with $N_t$ RHS elements and an RIS with $N_{RIS}$ elements.}
	\label{Fig:RHS_RIS_UE_location}
\end{figure}

The simulations {rely on} a configuration, where random initial values are assigned to $\mbf{F}$, $\mbf{M}$, and $\bols{\Theta}_{RIS}$. The iterative process of updating one variable, while holding the others constant is applied to all three variables. This iterative procedure continues until a specified stopping criterion is satisfied, as defined in {Algorithm}~\ref{Algo:AM_algorithm}. {The solutions to the three subproblems optimizing $\mbf{F}$, $\mbf{M}$, and $\bols{\Theta}_{RIS}$ are either near-optimal or locally optimal, as analyzed in the previous subsections. Therefore, the term `optimal' describes the best solution obtained within the constraints of a specific algorithm, but it does not necessarily represent the global optimum of the overall problem}. The  parameters used for the simulations are given in Table~\ref{Tab:Simulation Parameters}. The simulation results
are averaged over {100} independent channel realizations. The wireless channels are generated using the mmWave channel model~\eqref{Eq:mmwave_chan_model} discussed in Section~II. The flow chart of the joint sum-rate optimization algorithm is summarized in Fig.~\ref{Fig:Flow_chart_AM}. We compare the performance of the proposed method {to a range of} benchmark schemes: (a) an RHS system without an RIS, and (b) an RHS system with a randomly configured RIS. In the latter case, we consider a random selection of the passive RIS phase shift matrix $\bols{\Theta}_{RIS}$. For a fair comparison, we employ the same objective function and constraints across all benchmark schemes, using the sum-rate as the performance metric.\footnote{In the simulation, the UE locations are ($98.3$m, $27.8$m), ($99.8$m, $30.1$m), ($100.2$m, $30.7$m) and ($99$m, $32.9$m)}

The BS is equipped with $N_t= 64$ {elements} located at ($0$m, $0$m), and $4$ single-antenna UEs ($K = 4$) uniformly and randomly distributed in a circle centered at ($100$m, $30$m) with radius $10$ m. We assume that the RIS is deployed at (100m, 0m), and the UEs location is fixed once randomly generated {as shown in Fig.~\ref{Fig:RHS_RIS_UE_location}}. {The $N_{RF}$ RF chains are connected to the $N_t$ RHS elements through $N_{RF}$ feeds that convert the carrier frequency current into an electromagnetic wave. This wave propagates through the waveguide of the RHS and radiates the energy into the free space from the  $N_t$ RHS elements, as shown in Fig.~\ref{Fig:RHS_RIS_system}. }

{For the mmWave model, the channel gains $\mbf{\alpha}_{l_*}$ are generated independently, following the distribution $\mbf{\alpha}_{l_*} \thicksim \mathcal{CN}(0,10^{-0.1PL(d_i)})$. The quantity $PL(d_i)$ represents the path loss dependent on the distance  $d_i$ associated with the corresponding link. The modified path loss model described in~\cite{Hanzo_2017_IEEE_Comsurvey_mmwave} is considered, and the path loss in dB is given by the following equation
\begin{equation}
	PL(d_i) = a + 10b\;\mrm{log_{10}}(d_i)+ \kappa ,
\end{equation}
where $\kappa \in \mathcal{CN}(0,\sigma^2_\kappa)$.
The values of $a, b$ and $\sigma^2_\kappa$ are set to $a = 61.4$, $b = 2$ and $\sigma^2_\kappa = 5.8$ dB for the line-of-sight paths.
For the non-line-of-sight (NLOS) paths, the values are set to  $a = 72$, $b = 2.92$ and $\sigma^2_\kappa = 8.7$ dB. To evaluate the effectiveness of the RIS more accurately, we assume that each path of $\mbf{H}_d$ is a NLOS path that faces partial blockage from obstacles and experiences an additional penetration loss of 40 dB. The carrier frequency ($f_c$) is
set to $28$~GHz, the bandwidth (B) is set to $250$~MHz and thus
the noise power is $-174+10\;\mrm{log_{10}}$B = $-90$ dBm. For simplicity, we consider each UE to experience the same number of paths from both the BS and the RIS.  Moreover, we set the number of paths to
$L_i$ = 10, $i \in \{d, ru, bu\}$. The antenna spacing at the RHS and RIS is set to $d =\lambda/4$ and $\lambda/2$, respectively. }

The RIS is applied to provide {a} high-quality link between the BS and UEs, {where} we assume that the LOS component is {included in} the channel between {the} BS and RIS, and {in the} channel between {the} RIS and each UE. {Benefited from the directional reflections of the RIS, the BS-RIS-UE link is usually stronger than other multipaths as well as the degraded direct link between the BS and the UE}.
Five iterations of the proposed algorithm have been considered in this analysis. In each iteration, the three sub problems are solved alternatively.

\begin{figure}[hbt!]
	\centering
	\includegraphics[width=\linewidth]{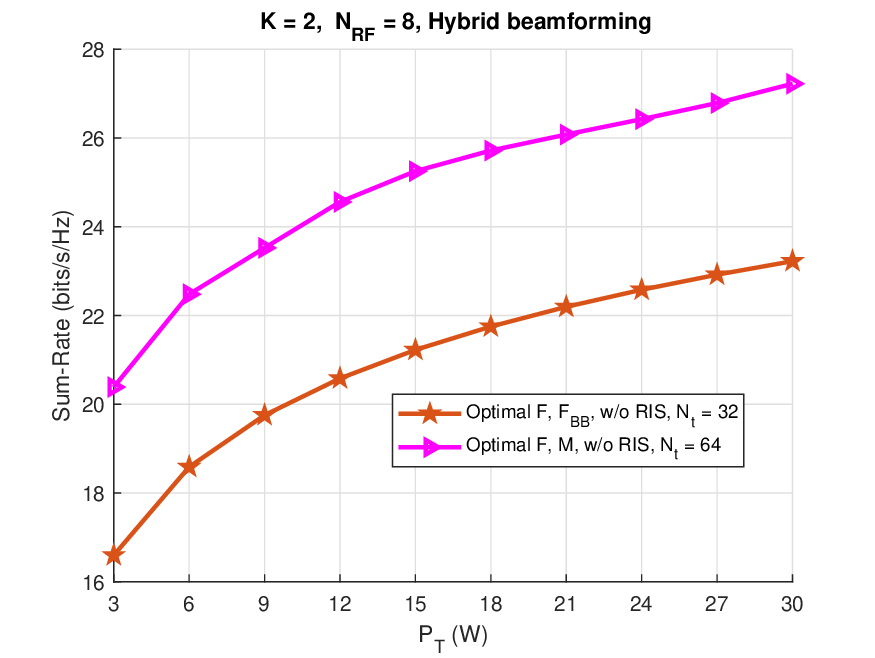}
	\caption{\footnotesize The sum-rate for hybrid beamforming using active phased arrays and holographic beamforming is compared across different values of $P_T$ for the case where $K=4$.}
	\label{Fig:RHS_RIS_HBF}
\end{figure}

{In Fig. \ref{Fig:RHS_RIS_HBF}, the sum-rate comparison for hybrid beamforming using active phased arrays and holographic beamforming without an RIS is presented across different  $P_T$ values for $K = 2$. The analog beamforming matrix using active phased arrays is represented by $\mbf{F}_{BB}$ with the distance between elements set to $\lambda/2$.  In contrast, for holographic beamforming, the distance between RHS elements is $\lambda/4$. This allows for twice the number of smaller RHS elements to be accommodated within the same space. It is evident that the sum-rate gain is consistently higher when using holographic beamforming compared to analog beamforming with active phased arrays.}

\begin{figure*}[hbt!] 
	\centering	
	\begin{subfigure}[]{0.33\linewidth}
		\centering
		\includegraphics[width=\linewidth]{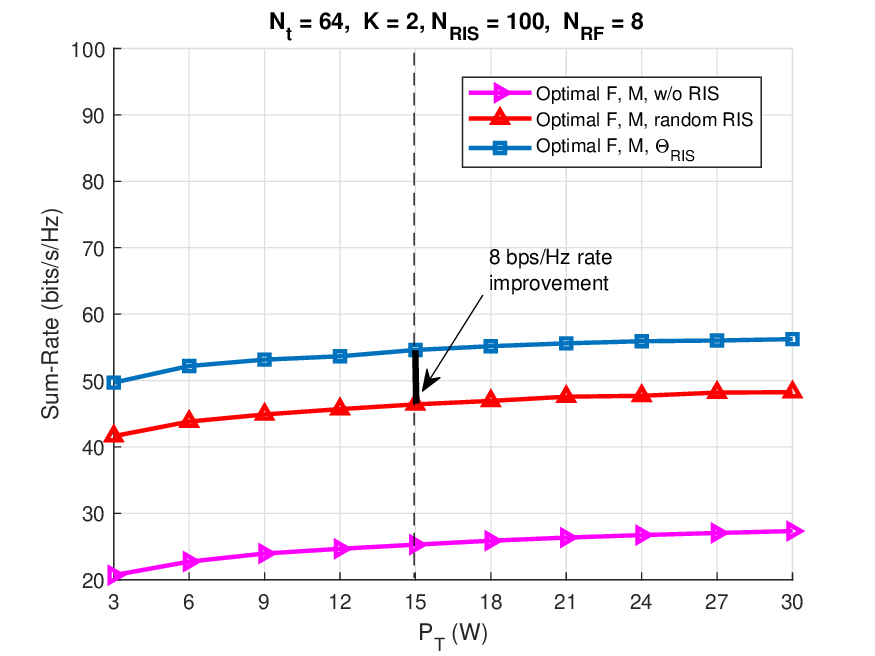}
		\caption{$K=2$}
		\label{Fig:RHS_RIS_compare1}	
	\end{subfigure}%
	\begin{subfigure}[]{0.33\linewidth}
		\centering	\includegraphics[width=\linewidth]{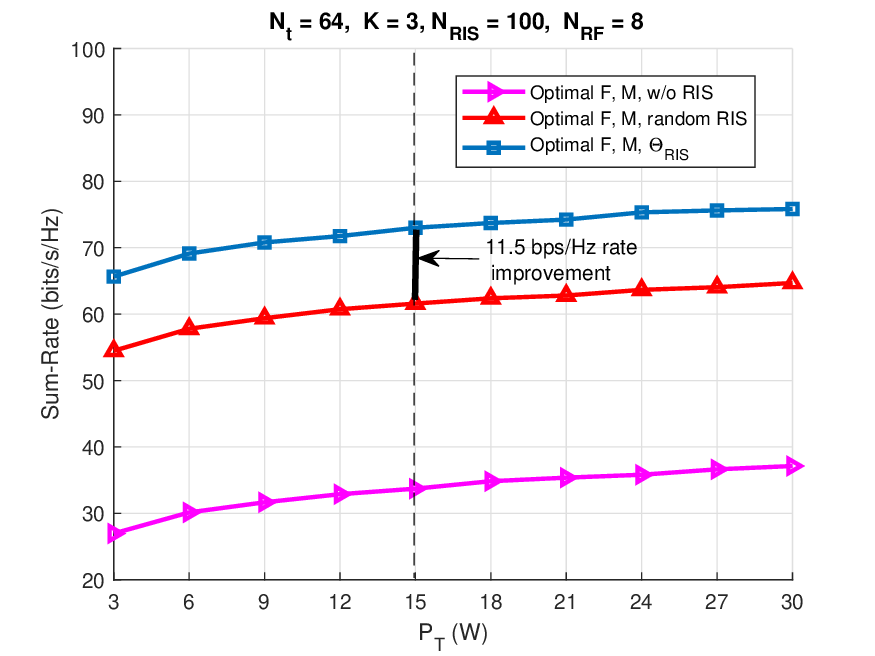}
		\caption{$K=3$}
		\label{Fig:RHS_RIS_compare2}
	\end{subfigure}%
	\begin{subfigure}[]{0.33\linewidth}
		\centering
		\includegraphics[width=\linewidth]{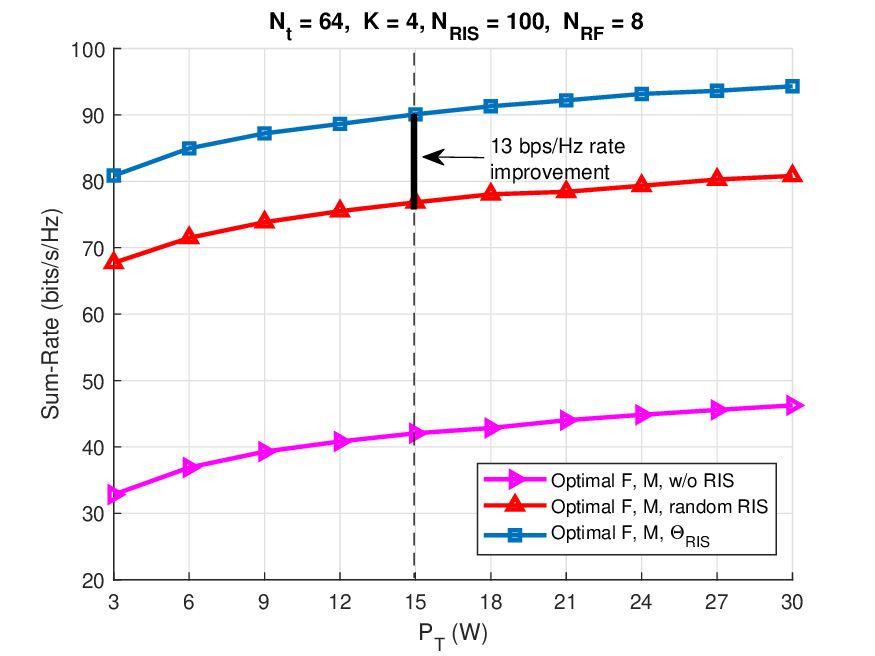}
		\caption{$K=4$}
		\label{Fig:RHS_RIS_compare3}
	\end{subfigure}%
	\caption{\footnotesize The sum-rate is compared across different values of $P_T$ for the case where $K=2,3$ and $4$, considering a given set of parameters including $N_t, N_{RF}$, and $N_{RIS}$ }
	\label{Fig:RHS_RIS_compare}
\end{figure*}

In Figs.~\ref{Fig:RHS_RIS_compare1}, \ref{Fig:RHS_RIS_compare2}, and \ref{Fig:RHS_RIS_compare3}, the sum-rate is illustrated for $K = 2, 3$ and $4$ {and for} different $P_T$ values.
It can be observed that when there is no RIS in the system, the performance gain in sum-rate is always less than {that of} the system {relying on} an RIS. If the phase shift matrix $\bols{\Theta_}{RIS}$ is not optimized, the performance gain achieved by deploying an RIS is insignificant, as expected.
For example, {observe} in Fig. \ref{Fig:RHS_RIS_compare3} {that} hybrid beamforming and {the} phase optimization scheme {applied by} the system
achieve about {13} bps/Hz improvement in sum-rate at $P_T=$ 15 W. It is observed that the proposed algorithm {relying on} an optimized phase shift matrix $\bols{\Theta}_{RIS}$ exhibits the highest sum-rate for all the values of $P_T$ across all methods. Clearly, for higher values of $P_T$ at the BS {we can have an increased sum-rate}, while keeping the remaining system parameters constant.

\begin{figure}[hbt!]
	\centering
	\includegraphics[width=\linewidth]{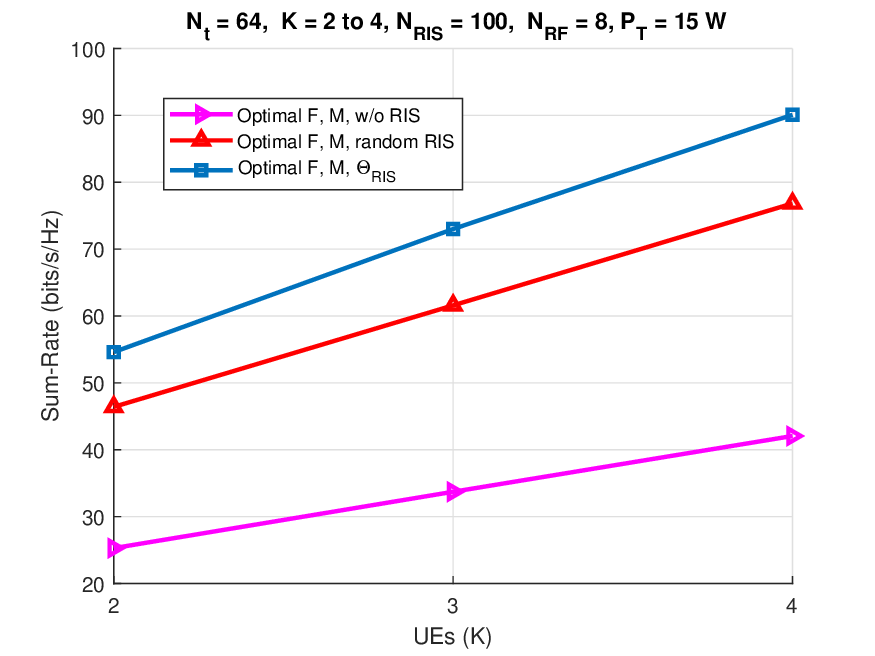}
	\caption{\footnotesize The sum-rate is compared across different UEs for a given set of parameters including  $N_t, N_{RF},  N_{RIS}$ and $P_T$.}
	\label{Fig:RHS_RIS_UE_compare}
\end{figure}

\begin{figure}[hbt!]
	\centering
	\includegraphics[width=\linewidth]{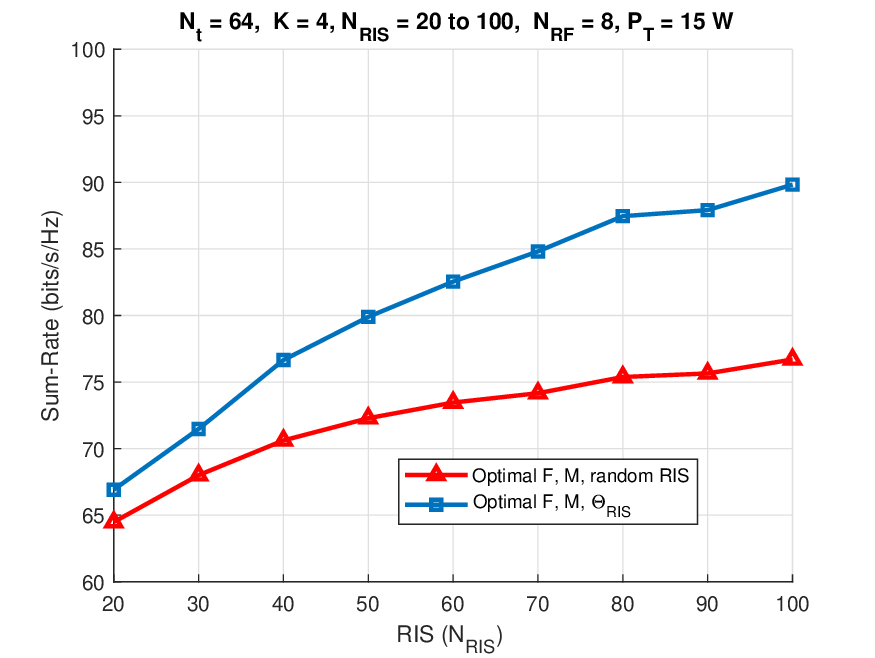}
	\caption{\footnotesize The sum-rate is compared across different $N_{RIS}$ for a given set of parameters including  $N_t, N_{RF},  K$ and $P_T$.}
	\label{Fig:RHS_RIS_RIS_compare}
\end{figure}

\begin{figure}[hbt!]
	\centering
	\includegraphics[width=\linewidth]{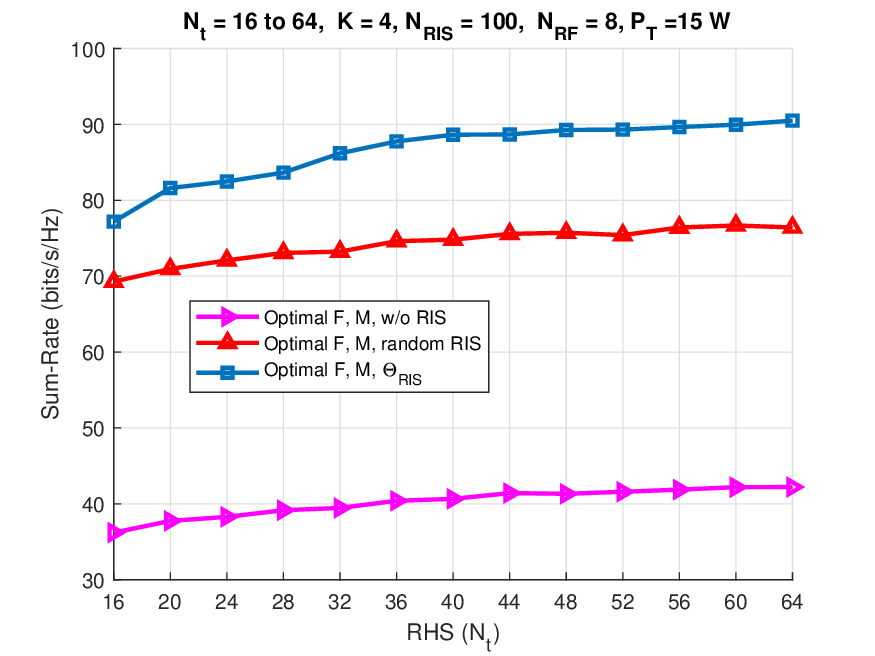}
	\caption{\footnotesize The sum-rate is compared across different $N_t$ RHS elements, for a given set of parameters including $N_{RF}$, $K$ and $P_T$.}
	\label{Fig:RHS_RIS_RHS_compare}
\end{figure}

In Fig. \ref{Fig:RHS_RIS_UE_compare} the sum-rate is illustrated for $K = 2, 3$ and $4$ {and for} a $P_T$ value of {15} W. Again, it can be observed that when there is no RIS in the system and  when using a random phase shift matrix, the sum-rate {gain} is always less than that of the system with an optimized RIS.
For example, in Fig. \ref{Fig:RHS_RIS_UE_compare}, the system achieves about {11.5 bps/Hz} improvement in sum-rate at $K=$ 3 {upon using phase optimization}. It is also observed that the proposed algorithm {having} an optimized phase shift matrix $\bols{\Theta}_{RIS}$ exhibits the highest sum-rate for all the values of $K$ {for} all methods. Clearly, for higher values of $K$ at the BS {we have an increased sum-rate}, while keeping the remaining system parameters constant.

In Fig. \ref{Fig:RHS_RIS_RIS_compare}, a comparison of the sum-rate is illustrated for $N_{RIS}$ ranging from 20 to {100} with a $P_T$ value of {15} W. Once again, it is evident that the sum-rate {gain} is consistently lower, when using a random phase shift matrix compared to a system with an optimized phase shift matrix $\bols{\Theta}_{RIS}$ {for} all values of $N_{RIS}$.
Clearly, for higher values of $N_{RIS}$ {we have an increased sum-rate}, while keeping the remaining system parameters constant.

In Fig. \ref{Fig:RHS_RIS_RHS_compare}, a comparison of the sum-rate is illustrated for RHS elements ranging from 16 to 64 with a $P_T$ value of 15 W. {With an optimized phase shift matrix, increasing the number of RHS elements from 16 to 64 results in a sum-rate improvement of 13 bps/Hz.} Once again, it can be observed that in the absence of an RIS and when using a random phase shift matrix, the {sum-rate gain} is consistently lower than that achieved by the system {having} an optimized phase shift matrix $\bols{\Theta}_{RIS}$ {for} all values of $N_t$. Clearly, for higher values of $N_t$, {we have an increased sum-rate}, while keeping the remaining system parameters constant.

\begin{figure}[hbt!]
	\centering
	\includegraphics[width=\linewidth]{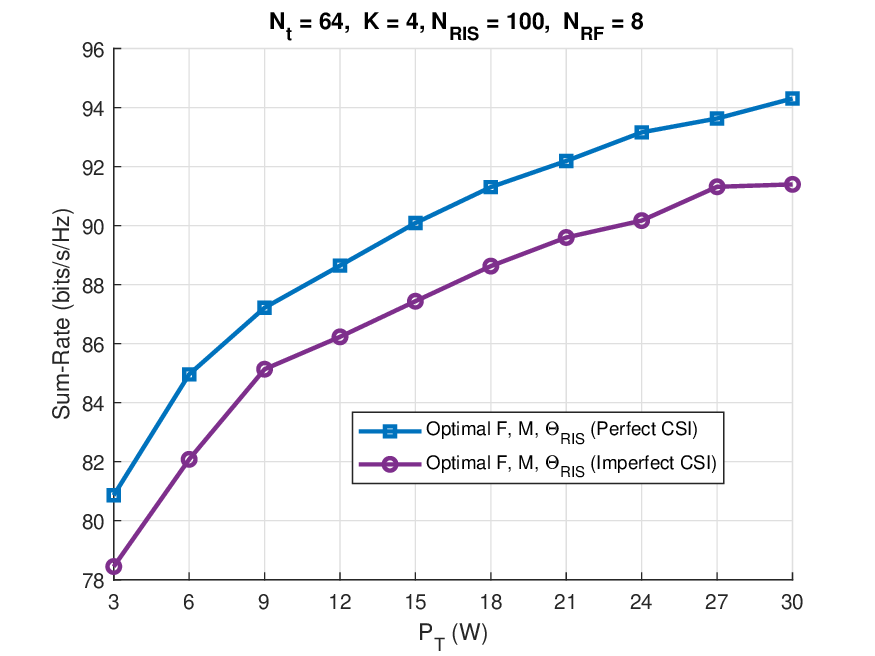}
	\caption{\footnotesize The sum-rate for both perfect and imperfect CSI is compared across different values of $P_T$ for the case where $K=4$.}
	\label{Fig:RHS_RIS_perf_imperf}
\end{figure}

{Fig.~\ref{Fig:RHS_RIS_perf_imperf} presents a comparison of the sum-rate for both perfect and imperfect CSI, evaluating the robustness of the proposed sum-rate maximization algorithm against channel estimation errors~\cite{Erik_2020_IEEETWC_RIS_weightedsumrate,Lingyang_2022_IEEETWC_RHS_wireless}. In the imperfect CSI case, we assume that the channel matrix from the RHS to the $k$-th UE $\mbf{H}_k$, lies within a ball of radius 0.1$\|\mbf{H}_k\|$ around the estimated channel matrix, $\hat{\mbf{H}}_k$. This is expressed as  $\mbf{H}_k = \{\hat{\mbf{H}}_k + \bols{\delta}_k \;| \;\|\bols{\delta}_k\| \leq 0.1 \|\mbf{H}_k\|\}$, where $\bols{\delta}_k$ represents the channel estimation error,  with its norm assumed to be bounded by 0.1$\|\mbf{H}_k\|$. A slight drop in the sum-rate can be observed, indicating the inherent robustness of the proposed algorithm against channel estimation errors.}

\begin{figure}[hbt!]
	\centering
	\includegraphics[width=\linewidth]{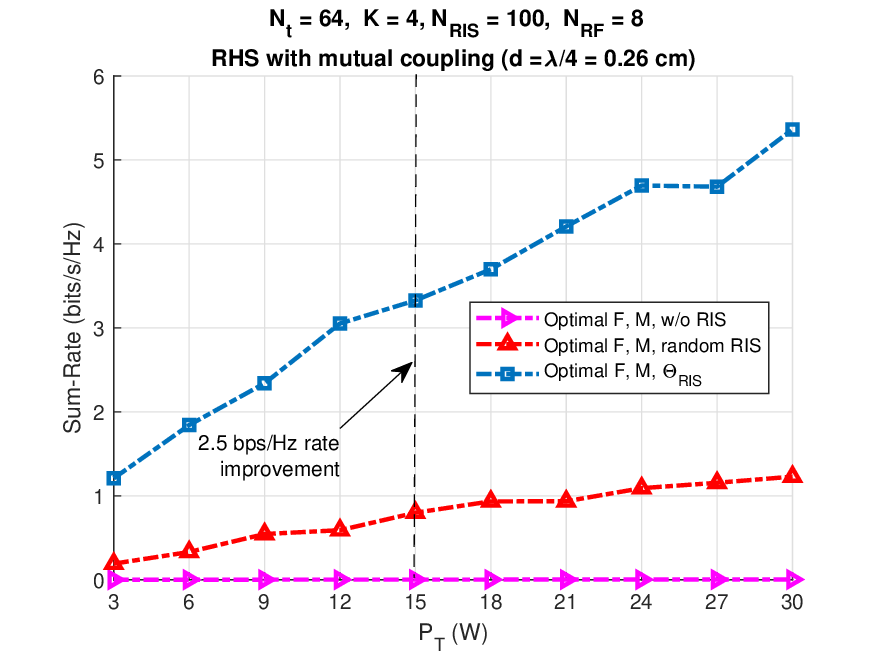}
	\caption{\footnotesize The sum-rate in the presence of mutual coupling, generated using~\eqref{Eq:MC_matrix} for a distance of $\lambda/4 = 0.26$ cm between the RHS elements, is compared across different values of $P_T$ for the case where $K=4$.}
	\label{Fig:RHS_RIS_mutual_coup}
\end{figure}

{In Fig. \ref{Fig:RHS_RIS_mutual_coup}, the sum-rate in the presence of mutual coupling is illustrated for $K = 4$ across different $P_T$ values. It can be observed that when the distance between the RHS elements is $\lambda/4$ and the system operates at mmWave frequencies, the high mutual coupling leads to a significant loss in sum-rate performance. However, the proposed algorithm, which relies on an optimized phase shift matrix $\bols{\Theta}_{RIS}$ consistently achieves the highest sum-rate for all $P_T$  values across all methods, even in the presence of mutual coupling. This demonstrates that our proposed method performs well in both scenarios, regardless whether mutual coupling is present or absent.}

	\section{Summary and Conclusion}
	
	We {proposed} a novel AM algorithm designed {for maximizing} the sum-rate of {a} mmWave multi-UE system. The proposed manifold optimization-based AM algorithm {leveraged} all the new wireless technologies like RHS and RIS in the face of {a} mmWave channel. The numerical results {revealed} a significant enhancement in the sum-rate when employing our proposed algorithm. {The sum-rate improvement attained ranges from 8 bps/Hz to 13 bps/Hz for $K$ values ranging from 2 to 4 and for various $P_T$ values, which is significant.}
	This confirms the effectiveness of our algorithm over existing methods in mmWave multi-UE systems.

\section{Acknowledgement}

\thanks{Pavan Kumar Gadamsetty and K.V.S. Hari  would like to acknowledge the support of the British Telecom India Research Centre (BTIRC), Indian Institute of Science, Bangalore.} \\
\thanks{L. Hanzo would like to acknowledge the financial support of the Engineering and Physical Sciences Research Council (EPSRC) projects under grant EP/Y037243/1, EP/W016605/1, EP/X01228X/1, EP/Y026721/1, EP/W032635/1 and EP/X04047X/1 as well as of the European Research Council’s Advanced Fellow Grant QuantCom (Grant No. 789028). }

\appendix

\section{Appendix}

\noindent {In the Dinkelbach-based method, $\mathbf{m}^{(0)}$ represents the randomly initialized value of $\mbf{m}$, $\mathbf{\Sigma}^{{{\mathcal{P}}_2}}$ is defined as the summation of $\mathbf{\Sigma}_k^{{{\mathcal{P}}_2}}$ for $k$ from 1 to $K$, and $\widetilde{\mathbf{\Sigma}}^{{{\mathcal{P}}_2}}$ is defined as the summation of $\widetilde{\mathbf{\Sigma}}_k^{{{\mathcal{P}}_2}}$ for $k$ from 1 to $K$. The optimal value of $\mbf{m}$ obtained after solving~\eqref{Eq:RHS_Solution} using Algorithm~\ref{Algo:Dinkelbach} is denoted by $\mathbf{m}^*$.
Dinkelbach method can
be convergent to the global optimal solution~\cite{Mahdi_2015_IEEEJSTSP_FPoptimizing}.}


\begin{algorithm}[] 
	\DontPrintSemicolon
	\SetAlgoLined
	\SetKwInput{kwInit}{Init}
	\SetKw{Kw}{Variables}
	\KwIn{$\mathbf{\Sigma }^{{\mathcal{P}}_2}$, $\widetilde {\mathbf{\Sigma }}^{{{\mathcal{P}}_2}}$, ${\zeta}$, $\mbf{m}^{(0)}$}  
	\KwOut{$\mbf{m}^{*}$}
	\vspace{1mm}
	Set $t=0$, $\lambda_{t}=0$ and  $\mbf{m}_{t} = \mbf{m}^{(0)}$  \;
	\vspace{1mm}

	\tbf{repeat}  \;
	\hspace{1.5mm}Find the optimal solution $\mbf{m}_{t}$ by solving problem~\eqref{Eq:RHS_Solution} using CVX. \;

	\hspace{1.5mm}Let $F^{}_{\lambda_t} = 2{{\mathbf{m}^{(0)}}^T}\Re \left( {{\mathbf{\Sigma }}^{{{\mathcal{P}}_2}}} \right){\mathbf{m}_{t}} - \lambda_t  {{\mathbf{m}}^{{T}}_{t}}\Re \left( {\widetilde{\mathbf{\Sigma }}^{{{\mathcal{P}}_2}}} \right){{\mathbf{m}}_{t}}$;   \;
	\hspace{1.5mm}Set $t = t + 1$; \;
	
	\hspace{1.5mm}Update $\lambda_{t} = {\frac{{2{{\mathbf{m}}^{{{(0)}^T}}}\Re \left( {{\mathbf{\Sigma }}^{{{\mathcal{P}}_2}}} \right){\mathbf{m}_{t}} - {{\mathbf{m}}^{{{(0)}^T}}}\Re \left( {{\mathbf{\Sigma }}^{{{\mathcal{P}}_2}}} \right){{\mathbf{m}}^{(0)}}}}{{{{\mathbf{m}}^T_{t}}\Re \left( {\widetilde {\mathbf{\Sigma }}^{{{\mathcal{P}}_2}}} \right){\mathbf{m}_{t}} + \sigma ^2}}}$; \;
	
	\tbf{until}  
	$F_{\lambda_t}\leq {\zeta} $  or maximum iterations reached  \;
	\tbf{return} $\mbf{m}^{*} = \mbf{m}_{t}$  \;
	
	\caption{{Dinkelbach-based solver for~\eqref{Eq:RHS_Solution}}}
	\label{Algo:Dinkelbach}
\end{algorithm} 



	\medskip
	\bibliography{References}
	\bibliographystyle{IEEEtran}
  \vspace{-20mm}
	\begin{IEEEbiography}[{\includegraphics[width=1in,height=1.25in,clip,keepaspectratio]{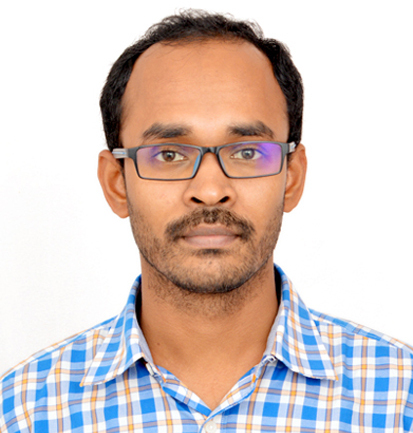}}]{Pavan Kumar Gadamsetty } (Student Member, IEEE) received the B.E. degree in Electronics and Communication engineering from Andhra University, Visakhapatnam, India, in 2011,
		the M.Tech. degree in Electrical engineering from the Indian Institute of Technology (IIT), Kanpur, in 2014. From August 2014 to December
		2018, he worked at Cisco Systems, Bangalore, India. He is currently pursuing a Ph.D. degree in Electrical Communication engineering from the Indian Institute of Science, Bangalore.
		His current research interest includes channel estimation techniques, precoding techniques for massive MIMO systems, and emerging technologies like RIS and RHS in wireless communications.
	\end{IEEEbiography}
	\vspace{-20mm}
	\begin{IEEEbiography}[{\includegraphics[width=1in,height=1.25in,clip,keepaspectratio]{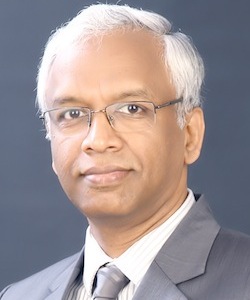}}]{K V S HARI } (Fellow, IEEE)  is a Professor in the Department of ECE, and  Director, Centre for Brain Research, Indian Institute of Science (IISc), Bangalore. He holds a PhD (Systems Science) from U C San Diego and has been a visiting faculty at Stanford University and Affiliate Professor at KTH-Royal Institute of Technology, Stockholm. His research interests are in Signal Processing and Deep Learning with applications to 5G wireless communications, indoor positioning, dual function radar and communication systems, autonomous navigation, neuroscience, and affordable MRI systems. He is a co-author of the IEEE 802.16 standard on wireless channel models and also served as Chair, Standardisation Committee, Telecom Standards Development Society, India. He was leading the British Telecom India Research Centre (BTIRC) at IISc.   He was an Editor of EURASIP’s Signal Processing and is currently the Editor-in-Chief (Electrical Sciences) of Sadhana, the journal of the Indian Academy of Sciences published by Springer. He is a Fellow of the Indian National Academy of Engineering, Indian National Science Academy and IEEE. He was on the Board of Governors, IEEE Signal Processing Society as VP-Membership. More details at \url{http://ece.iisc.ac.in/~hari}.
		
	\end{IEEEbiography}
	\vspace{-20mm}
	\begin{IEEEbiography}[{\includegraphics[width=1in,height=1.25in,clip,keepaspectratio]{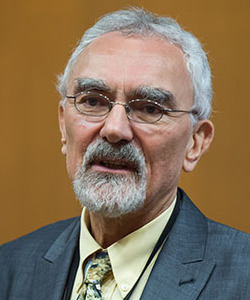}}]{Lajos Hanzo} is a Fellow of the Royal Academy of Engineering, FIEEE, FIET, Fellow of EURASIP and a Foreign Member of the Hungarian Academy of Sciences. He coauthored 2000+ contributions at IEEE Xplore and 19 Wiley-IEEE Press monographs. He was bestowed upon the IEEE Eric Sumner Technical Field Award.  More details at (\url{http://www-mobile.ecs.soton.ac.uk}, \url{https://en.wikipedia.org/wiki/Lajos_Hanzo})
	\end{IEEEbiography}

\end{document}